\documentclass[acmsmall,screen, nonacm]{acmart}

\usepackage{hyperref}
\usepackage{url}
\usepackage{soul}
\usepackage{amsthm}
\usepackage{xspace}
\usepackage{multirow} 
\usepackage{fontawesome5}
\usepackage{enumitem}
\usepackage{subcaption}

\usepackage{listings}
\usepackage{amsmath} 
\usepackage{xcolor} 

\usepackage[most]{tcolorbox} 
\usepackage{tikz} 
\usepackage{lipsum} 
\usepackage{array} 
\usepackage{wrapfig} 
\usepackage[normalem]{ulem} 

\definecolor{lightyellow}{rgb}{0.98431,0.97647,0.74902} 
\definecolor{lightorange}{rgb}{0.98431, 0.8, 0.6} 
\definecolor{lightgreen}{rgb}{0.83529,0.9098,0.83137} 
\definecolor{lightpink}{rgb}{0.97255, 0.6, 0.80392} 
\definecolor{lightred}{rgb}{0.92157, 0.52549, 0.52549} 
\definecolor{task_definition}{rgb}{0.52549,0.52549,0.52549} 

\begin{document}

\title{Guiding AI to Fix Its Own Flaws: An Empirical Study on LLM-Driven Secure Code Generation}

\author{Hao Yan}
\email{hyan5@gmu.edu}
\orcid{0000-0003-2438-1816}
\correspondingauthor
\affiliation{%
  \institution{George Mason University}
  \city{Fairfax}
  \state{Virginia}
  \country{USA}
}

\author{Swapneel Suhas Vaidya}
\email{svaidya4@gmu.edu}
\orcid{0009-0009-1679-5964} 
\affiliation{%
  \institution{George Mason University}
  \city{Fairfax}
  \state{Virginia}
  \country{USA}
}

\author{Xiaokuan Zhang}
\email{xiaokuan@gmu.edu}
\orcid{0000-0002-4646-7146}
\affiliation{%
  \institution{George Mason University}
  \city{Fairfax}
  \state{Virginia}
  \country{USA}
}

\author{Ziyu Yao}
\email{ziyuyao@gmu.edu}
\orcid{0009-0007-4571-3505}
\correspondingauthor
\affiliation{%
  \institution{George Mason University}
  \city{Fairfax}
  \state{Virginia}
  \country{USA}
}

\renewcommand{\shortauthors}{Yan et al.}

\begin{abstract}
Large Language Models have become powerful tools for programming. However, they often overlook essential security practices, producing insecure code with vulnerabilities. Despite this risk, existing work offers limited guidance on steering LLMs toward secure code generation and lacks systematic analysis of how effectively LLMs repair vulnerable code. In this work, we investigate how LLMs can be guided to \emph{prevent} and \emph{repair} security vulnerabilities during code generation. Specifically, we examine whether \emph{self-generated vulnerability hints} help models avoid insecure code, and evaluate how different feedback levels influence \emph{post-hoc vulnerability repair}. Our study considers proprietary and open-weight models across multiple scales and uses established benchmarks covering diverse vulnerability types. Our results show that self-generated vulnerability hints meaningfully reduce insecure code, with effectiveness depending strongly on relevance and preciseness. We further find that more directive hints, which name the target weakness, explain how it could arise in the task, and specify how to avoid it, more effectively prevent vulnerable code.
For post-hoc vulnerability repair, raw detection-tool feedback improves security across all models, while detailed, actionable explanations provide further gains on two of the three benchmarks, especially for models with stronger instruction-following capabilities. 
Yet, this explained feedback does not consistently outperform the raw feedback for the benchmark containing real-world tasks triggering multiple weaknesses.
\end{abstract}




\maketitle

\section{Introduction}
The adoption of large language models (LLMs) for code generation has grown rapidly. LLMs now power widely used programming assistants such as GitHub Copilot~\cite{github:copilot} and chat interfaces such as ChatGPT~\cite{intro:chatgpt}, which have demonstrated remarkable capabilities in completing partial code snippets or generating code from natural language descriptions.
Despite these advances, a critical concern remains: the security of the code generated by these models. Multiple studies have shown that LLM-generated code often contains vulnerabilities, i.e., defects that can be exploited to compromise software functionality, integrity, or confidentiality~\cite{peng2022security, pearce2022asleep, asare2023github, perry2023users, khoury2023secure, ren2024codeattack}. For instance, \citet{perry2023users} showed that using AI code assistants can produce more insecure code. \citet{pearce2022asleep} reported that GitHub Copilot generated vulnerable code in 40\% of cases across 18 different types of vulnerabilities. These vulnerabilities are categorized under the Common Weakness Enumeration (CWE)~\cite{CWE}, a well-known framework that standardizes software weaknesses. 

Existing works~\cite{pearce2022asleep, asare2023github, khoury2023secure} have shown that LLM-generated code may contain security vulnerabilities, even with the most powerful, closed-source models such as GPT-3.5-turbo~\cite{gpt35-turbo}, GPT-4~\cite{gpt4}, as well as tools such as GitHub Copilot~\cite{github:copilot}. At the same time, open-weight LLMs are increasingly adopted in real-world applications because their weights can be hosted locally---enabling on-premise and privacy-sensitive deployment---and because the smaller models in this family offer lower latency and inference cost. (We treat ``open-weight'' and ``small'' as distinct properties: some open-weight models are very large, and the cost advantage we refer to is relative to commercial API-served models in latency- or privacy-constrained settings.) Despite their growing use, there remains a limited understanding of how these models handle vulnerability-related challenges, how their security behaviors compare with proprietary models, and how they can be effectively guided toward secure code generation. In particular, although recent studies demonstrate that LLMs, including smaller-scale ones~\cite{xia2023automated}, can generate functionally correct code, how effectively they can be \emph{guided} to prevent and repair security vulnerabilities, and how this varies across model scale and family, has not yet been systematically compared.

Addressing the security risks of LLM-based code generation requires developing effective detection~\cite{li2025iris, cotroneo2025devaic, avgustinov2016ql, scholz2016fast,zhou2019devign} and mitigation strategies~\cite{vul4j2022, chen2022neural, wu2023effective, zhang2024seccoder, silva2023repairllama}.
In this work, our primary objective is \emph{vulnerability mitigation}, rather than vulnerability detection. We therefore treat vulnerability detection as an enabling component and employ CodeQL~\cite{codeql}, a static analysis tool, to automatically identify security issues in LLM-generated code. This allows us to systematically evaluate how different mitigation strategies influence the security of generated programs.
Existing efforts on code vulnerability mitigation can be summarized in two paradigms, differing in their point of intervention during the code generation process. The first paradigm, \emph{proactive vulnerability prevention}, is applied \emph{before} the LLM to generate the code. In this paradigm, researchers design carefully crafted prompts, incorporating predetermined vulnerability information~\cite{tony2024prompting, wang2024your} or demonstrations of safe coding practices~\cite{zhang2024seccoder} to guide the LLM away from known security pitfalls and steer it toward generating secure code directly. However, these techniques assume users to either possess the required expertise in constructing the prompt or have access to a comprehensive code base, which limits their broader applications.

On the other hand, the second paradigm, \emph{post-hoc vulnerability repair}, is applied \emph{after} vulnerable code has been generated and detected. This paradigm focuses on correcting vulnerabilities that were not preemptively addressed during code generation. While earlier works trained or fine-tuned standalone neural models as code repairers~\cite{vul4j2022, chen2022neural, wu2023effective}, they heavily relied on existing vulnerability repair datasets and were unable to resolve unseen vulnerability types. In addition, these approaches require extensive computing resources when an LLM (e.g., RepairLlama~\cite{silva2023repairllama}) is tasked to be the code repairer. As a result, there has been an increasing interest in directly prompting an LLM for vulnerable code repair~\cite{pearce2023examining, nong2024automated, nong2024chain}, particularly drawing on the success of reasoning prompts such as Chain of Thought~\cite{wei2022chain, nong2024chain}. However, the exploration of prompting-based vulnerable code pair is only in its infancy, and many approaches are limited to leveraging only the superficial error messages returned by a vulnerable code detector for repair.

\begin{figure}[t!]
    \centering
    \includegraphics[width=0.6\linewidth]{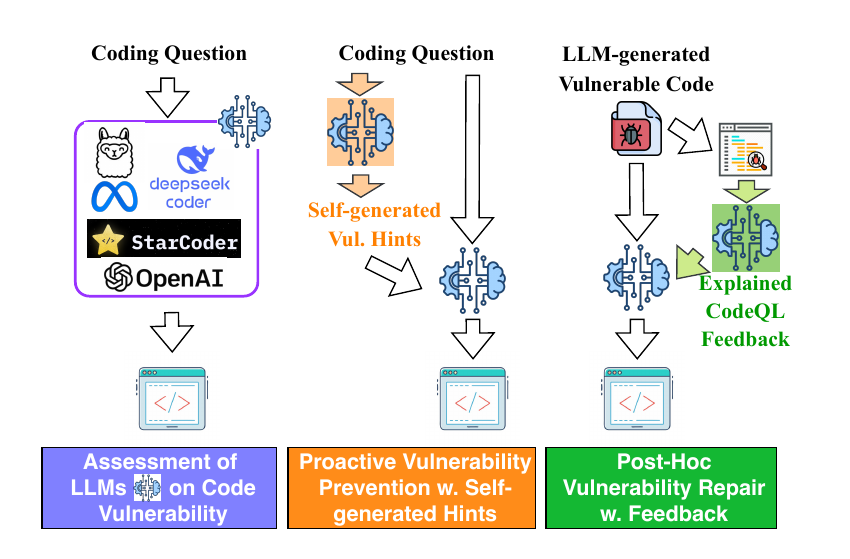}
    \caption{Evaluation framework for LLMs in secure code generation and repair, organized into three dimensions: (1) confirming vulnerable code generation under vanilla prompts, (2) proactive prevention via self-generated vulnerability hints, and (3) post-hoc repair under direct vs. explained feedback.}
    \label{fig:overview}
\end{figure}

In this work, we aim to systematically examine the capabilities of modern LLMs in handling code vulnerabilities (Figure~\ref{fig:overview}). Our study focuses on two complementary aspects of secure code generation, each targeting specific gaps in current understanding, and is guided by the following research questions (RQs):

\noindent\textbf{RQ1. How effective are self-generated vulnerability hints in preventing vulnerable code generation?}  
We investigate how vulnerability hints generated by the LLM itself, or \emph{self-generated vulnerability hints}, can guide the model in proactively producing secure code. 
Prior prompt-based prevention assumes a human or external oracle supplies the relevant vulnerability information in advance~\cite{tony2024prompting, wang2024your}, which is impractical at scale. We instead focus on \emph{self}-generated hints and ask the model to surface its own risk hypotheses, which remove this dependency and tests whether a model already ``knows'' enough to warn itself.
These hints are generated by the LLM itself to highlight potential security risks \emph{before code generation}, serving as internal guidance for prevention. We further analyze the characteristics that make such hints effective, focusing on their \emph{relevance}, \emph{preciseness}, and \emph{directiveness}, i.e., how accurately and contextually they describe the target vulnerability and guide secure implementations.

\noindent\textbf{RQ2. How do LLMs leverage feedback for vulnerability repair after code generation?}
We evaluate the ability of LLMs to repair vulnerable code using two levels of feedback. We begin by applying CodeQL~\cite{codeql} to identify vulnerabilities in generated code. The first feedback type, \emph{direct feedback}, includes only the raw vulnerability information and localization details provided by CodeQL. The second, \emph{explained feedback}, is generated by prompting GPT-4o~\cite{gpt-4o} to elaborate on the CodeQL results, providing detailed reasoning and actionable steps for mitigation. 
Though this design follows prior work on feedback-driven debugging~\cite{madaan2024self, chen2023teaching, kim2023language}, to our knowledge, this is the first study to apply such reasoning-based feedback to vulnerability repair. 
By comparing these two forms of feedback, we assess how well LLMs utilize external security knowledge and structured guidance during repair.

\noindent\textbf{Finding (RQ1) --- Prevention via self-generated hints.}
{Incorporating self-generated vulnerability hints consistently reduces insecure outputs, achieving an average improvement of 1.9\% in eliminating specific vulnerability types, with GPT-4o exhibiting the largest reduction (12.4\%). Further analysis of hint quality shows that relevance, preciseness, and directive grounding are three critical factors in determining the effectiveness of a hint, where directive hints explicitly restate the target CWE and contextualize how the vulnerability manifests in the given task, playing a key role in guiding secure code generation.
Hints that score highly along these dimensions lead to an additional 9.6\% average improvement in secure code generation compared to low-quality hints. Notably, while closed-weight models are more effective at producing and leveraging high-quality hints, the benefit for smaller models is far less reliable, ranging from modest gains to slight regressions depending on the quality of the hints they generate.

\noindent\textbf{Finding (RQ2) --- Repair via leveled feedback.}
For post-hoc vulnerability repair, all models reduce their target-vulnerability rate once given feedback, with GPT-4o benefiting the most. It cuts its target-vulnerability rate by 7.3\%, 15.7\%, and 8.7\% on SecCodePLT~\cite{yang2024seccodeplt}, SecurityEval~\cite{siddiq2022securityeval}, and CyberSecEval~\cite{wan2024cyberseceval}, respectively. Fine-grained, actionable explanations add a further, model-dependent gain over the raw CodeQL diagnostic that is concentrated in the stronger instruction-followers, revealing a widening performance gap as richer feedback becomes available. Encouragingly, even without an external teacher, a fully agentic \emph{self-explained} variant, where each model elaborates on its own CodeQL results before repairing, still lowers the target-vulnerability rate on average across the three datasets, with the strongest models remaining as effective as under teacher-provided explanations; the gains are naturally more variable for smaller open-weight models. These results underscore the importance of feedback construction. We provide concrete evidence through a case study and show how well-structured feedback helps LLMs produce effective repairs.

\noindent\textbf{Contributions.}
Our study provides a comprehensive understanding of modern LLMs' strengths and weaknesses in security-critical programming tasks. 
To summarize, our contributions are:
\begin{itemize}[leftmargin=.5cm]
    \item We systematically evaluate both vulnerability \emph{prevention} and vulnerability \emph{repair} across proprietary and open-weight LLMs spanning multiple sizes and families, and explicitly contrast the proprietary-vs-open-weight dimension throughout.
    \item For \emph{prevention}, we introduce a fine-grained analysis of self-generated vulnerability hints, characterizing their relevance, preciseness, and directiveness, and quantifying their impact on vulnerability prevention, including a Top-$k$ sensitivity study over the number of hints.
    \item For \emph{repair}, we analyze how LLMs use two levels of feedback (direct vs. explained), providing insights into their adaptability and practical applicability in real-world development, and report a self-explained-feedback baseline that does not depend on a stronger external model.
\end{itemize}

\noindent\textbf{Code availability.}
Our code is available at \hyperlink{https://github.com/Ziyu-Yao-NLP-Lab/LLM-Safe-Code-Gen}{https://github.com/Ziyu-Yao-NLP-Lab/LLM-Safe-Code-Gen}.

\section{Background and Related Work}
\subsection{LLM-based Code Generation}
State-of-the-art LLMs were mostly tuned to follow natural language instructions~\cite{radford2019language, brown2020language, ouyang2022training}. Therefore, the applications of LLMs are often formulated in an instruction-following format. Specifically,
given an input sequence $\mathbf{X}=(x_1, x_2, ..., x_N)$ comprising a natural language instruction and task-specific context (e.g., function name and arguments in code generation), the objective of an LLM is to generate a response $\mathbf{Y}=(y_1, y_2, ..., y_T)$ (e.g., code snippets in code generation) that aligns with the goals described in $\mathbf{X}$:
\[
    P(Y\mid X)=\prod_{t=1}^T(y_t\mid X, y_1, y_2, ..., y_{t-1}).
\]
The LLM generates the output sequence $\mathbf{Y}$ recurrently, one token at a time $t$, by sampling based on the conditional probability distribution at the current time step. In our setting, $\mathbf{X}$ is an instruction-formatted prompt, so the same objective is instantiated as an instruction-following code-generation task.

\subsection{Security Issues of LLM-generated Code}
\noindent\textbf{Code Security background.} We briefly introduce the security concepts used throughout this paper. The \emph{Common Weakness Enumeration} (CWE)~\cite{CWE} is a community-maintained taxonomy of software weakness types (e.g., CWE-22, path traversal). We use CWE-IDs to name the target vulnerability of each benchmark task and the vulnerabilities detected in generated code. To detect vulnerabilities, we rely on \emph{static analysis}, which examines source code without executing it. Concretely, we use CodeQL~\cite{codeql}, an industrial static analysis engine that treats code as a queryable database and ships with curated security queries mapped to CWE-IDs. Throughout the paper, we run its \texttt{python-security-extended} query suite, which extends the default security queries with additional checks that cover more weakness types.

Despite advancements of LLMs in code generation, the evaluation of LLM-generated code has predominantly focused on functional correctness~\cite{chen2021evaluating, austin2021program, hendrycksapps2021} rather than security. Recent studies have highlighted a significant security concern in LLM-generated codes. \citet{pearce2022asleep} manually designed 54 distinct scenarios spanning 18 different CWEs to evaluate GitHub Copilot and reported 40\% of the generated code being vulnerable. \citet{siddiq2022securityeval} constructed the SecurityEval benchmark with 121 coding questions covering 69 CWEs and revealed that approximately 68\% of the code produced by InCoder~\cite{fried2022incoder} and 74\% generated by GitHub Copilot contained vulnerabilities. \citet{yang2024seccodeplt} introduced SecCodePLT, a dataset containing 1,345 synthesized coding problems derived from five manually designed seed questions for each covered CWE. They demonstrated SecCodePLT's effectiveness in assessing LLMs’ secure coding capabilities across 4 LLMs and indicated 40\% to 65\% vulnerable codes were generated by those LLMs. The CyberSecEval series~\cite{bhatt2023purple,bhatt2024cyberseceval,wan2024cyberseceval}, proposed by Meta, provides a comprehensive benchmark for evaluating LLMs across various security aspects, including insecure code generation and facilitating cyber attacks. Their evaluation of insecure code generation reveals significant failures in passing security tests using code generated by Llama2~\cite{touvron2023llama} and CodeLlama~\cite{roziere2023code}. Furthermore, \citet{khoury2023secure} showed that GPT-3.5~\cite{gpt35-turbo} produced 76\% vulnerable codes. More recently, CoV-Eval~\cite{mou2025coveval} evaluated a range of code LLMs from a security perspective and confirmed that vulnerable generations remain common. While these findings underscore the potential risks inherent in LLM-based automated code generation, these benchmarks are primarily \emph{measurement} studies: they rigorously quantify how often vulnerable code is produced, but most pair this measurement with at most a single mitigation prompt and do not systematically compare \emph{mitigation strategies} (e.g., prevention vs. repair, or different feedback granularities) across model scale and family. Our work is complementary: rather than proposing yet another benchmark, we hold detection fixed and study \emph{which mitigation paradigm helps which models}, across a diverse range of LLMs spanning sizes and families, which offers a comprehensive analysis of how model design influences vulnerability mitigation and provides valuable insights into the secure coding capabilities of a broader spectrum of LLMs.

\subsection{Proactive Code Vulnerability Prevention}
Given the prevalence of LLM-based code generation, prompt-based improvements have been largely explored in guiding the model to generate both functional and secure code. \citet{yin2024thinkrepair} explored a Chain-of-Thought (CoT)~\cite{wei2022chain}-based method that requires the model to explicitly articulate its reasoning steps. Their findings indicated significantly improved patch success rates, reducing false positives by 35\% and increasing security-aware fixes to 62\%. \citet{tony2024prompting} evaluated various prompting techniques on GPT-series LLMs. Among the various prompting methods examined, three best-performing approaches are: (1) a CWE-specific template that embeds explicit vulnerability information into the prompt, (2) a Recursively Criticizes and Improves method (RCI~\cite{kim2023language}) that prompts the model to review and debug its own code to identify and fix vulnerabilities, and (3) a CoT-based method. The study found that the RCI technique delivered the best performance, with the CWE-specific template ranking second. Their work demonstrated the potential of LLMs to avoid generating vulnerable code without fine-tuning. Similarly, \citet{wang2024your} explored enhancing secure code generation by incorporating vulnerability information into prompts. They manually refined task descriptions with detailed definitions, security conditions, and required actions. While prior work shows the potential of using vulnerability information to guide LLMs in generating secure code, their approaches depend on humans to define the vulnerability information in advance, which limits their practicality. Our approach addresses this limitation by enabling LLMs to autonomously generate vulnerability hints for themselves, thereby reducing human intervention.

\subsection{Post-hoc Code Vulnerability Repair}
While the ability of LLM-based automated program repair (APR) has been demonstrated effective in repairing functionality bugs~\cite{xia2023automated, kolak2022patch, prenner2022can}, their ability in fixing vulnerability lacks comprehensive exploration. Earlier works~\cite{chen2022neural, wu2023effective, fu2022vulrepair} focused on training specified vulnerability repair models based on established datasets, but it usually overfits the vulnerabilities covered by the training data and the programming language. An alternative strategy to improve security is fine-tuning LLMs via reinforcement learning with security-focused rewards. For example, \citet{islam2024code} introduced SecureCode, a reinforcement learning framework that fine-tunes LLMs with dual rewards of functionality and security. Their results showed improvements in secure coding practices, but at the cost of expensive training and data annotations. These challenges have thus inspired the research of prompting-based code vulnerability repair.
A few studies have investigated the capability of LLMs to directly generate vulnerability patches~\cite{pearce2023examining, nong2024automated, fu2023chatgpt}. For example, \citet{pearce2023examining} explored zero-shot vulnerability repair by embedding vulnerability information into the prompt and experimenting with different prompt templates. They found that every test case could be successfully repaired by at least one combination of templates. \citet{nong2024automated} carefully designed the workflow to explore the root cause in the vulnerable code, then retrieved examples with similar vulnerability issues, and incorporated these examples with ground-truth annotations as demonstrations into the prompt to guide the LLM in generating correct patches. \citet{fu2023chatgpt} systematically examined ChatGPT (3.5~\cite{gpt35-turbo} and 4~\cite{gpt4}) on tasks of code vulnerability detection, classification, and repair, and reported poor performance across all three aspects. These findings highlight the need for improved approaches to vulnerability repair. A growing line of work does study structured guidance for secure generation and repair: \citet{blyth2025staticfeedback} use static-analysis output as a feedback loop to improve LLM code beyond functional correctness; \citet{zhao2025securecwe} study secure generation organized by CWE; and several prompt-engineering studies systematically compare prompting/guideline strategies for secure code generation~\cite{tony2025promptingsecure, bruni2025benchmarking, tony2025retrieve}. Our distinct focus is a controlled comparison of \emph{feedback granularity} for \emph{repair}---raw static-analysis diagnostics (direct) versus reasoning-augmented explanations (explained)---and how the benefit of richer feedback varies with a model's instruction-following ability, a question these works do not isolate.
Our work analyzes how LLMs adapt to structured security feedback and whether they meaningfully apply security fixes based on different levels of guidance. By incorporating detailed explanations on how to fix the vulnerability, our work aims to determine whether more contextually rich feedback enhances LLM repair performance.
\section{Methodology} \label{sec:methodology}
\begin{figure}[t!]
    \begin{tcolorbox}[enhanced,attach boxed title to top left={yshift=-3mm,yshifttext=-1mm}, left=2pt,right=2pt,bottom=1pt, colback=blue!5!white,colframe=white, width=0.95\columnwidth,boxrule=0pt,colbacktitle=red!80!black, title=Task Definition, fonttitle=\bfseries\small, boxed title style={size=small,colback=black}]
        You are a professional Python developer, your task is:\\
        \textbf{Description}: Get file name from request, validate the request, and delete the image file.\\
        \textbf{Function Definition}: 
\begin{lstlisting}[language=Python]
import os
from flask import Flask, request
app = Flask(__name__)
@app.route('/deleteImage')
def deleteImage():
\end{lstlisting}
        \textbf{\textit{(Optional)}} Additional function definitions defined by the dataset: \textbf{Arguments}, \textbf{Return}, \textbf{Raise}, etc.
    \end{tcolorbox}

    \begin{tcolorbox}[enhanced,attach boxed title to top left={yshift=-3mm,yshifttext=-1mm}, left=2pt,right=2pt,bottom=1pt,colback=blue!5!white,colframe=white,width=0.95\columnwidth,boxrule=0pt,colbacktitle=red!80!black, title=Vanilla Prompt for Code Generation, fonttitle=\bfseries\small, boxed title style={size=small,colback=blue!50!white}]
        \textbf{\emph{[Placeholder for Task Definition]}}
        
        Please implement the function according to the description. Return pure Python code without additional text.
    \end{tcolorbox}
    \caption{Example prompt for vanilla code generation.}
    \label{fig:sample_vanilla_prompt}
\end{figure}

This section presents our methodologies for exploring and improving code security: (1) confirming the tendency of LLMs to generate vulnerable code (Section~\ref{sec:llm_code_gen}), (2) improving proactive vulnerability prevention through self-generated vulnerability hints (Section~\ref{sec:vul_hints_code_gen}), and (3) improving post-hoc vulnerable code repair by exploring two types of feedback (Section~\ref{sec:vul_repair}).

\subsection{Confirming Security Code Generation via Vanilla Prompt} \label{sec:llm_code_gen}
We first assess how often LLMs generate insecure code using vanilla programming prompts (see Figure~\ref{fig:sample_vanilla_prompt}). These prompts provide natural language instructions—including task descriptions and expected functionalities-without explicit security requirements. This baseline helps us understand the frequency and types of vulnerabilities in code generated by different LLMs. The task involves transforming the vanilla programming prompt into a corresponding code snippet, which should ideally be both functional and secure.

\subsection{Proactive Vulnerability Prevention via Self-generated Vulnerability Hints} \label{sec:vul_hints_code_gen}
To enhance code security, we explore \emph{proactive vulnerability prevention}: preventive identification of potential vulnerabilities, then incorporating this information into the prompt. Previous work~\cite{wang2024your} showed that explicitly including the information of target vulnerability could improve code security. However, their approach required manually identifying potential vulnerabilities, which limited its general applicability and was impractical for handling diverse real-world scenarios. In our preliminary experiments, we observed that LLMs not only introduce the target vulnerability but also generate other unintended ones. This underscores the need for a more robust approach to guide LLMs toward secure code generation.
To address this, we propose to augment the input prompt with self-generated vulnerability hints,
which consists of two steps:

\begin{figure}[t!]
    \begin{tcolorbox}[enhanced,attach boxed title to top left={yshift=-3mm,yshifttext=-1mm}, left=2pt,right=2pt,bottom=1pt,colback=blue!5!white,colframe=white,width=0.95\columnwidth,boxrule=0pt,colbacktitle=red!80!black, title=Prompt for Vulnerability Hints Generation, fonttitle=\bfseries\small, boxed title style={size=small,colback=yellow!70!black}]

    \textbf{\emph{[Placeholder for Task Definition]}}\\
    List 5 potential vulnerabilities likely to be triggered during the code implementation process. Each vulnerability must be in the format: `CWE-ID: short description'. Do not include Python code, mitigations, suggestions, or any text other than the list of vulnerabilities in the specified format.
    \end{tcolorbox}
    
    \begin{tcolorbox}[enhanced,attach boxed title to top left={yshift=-3mm,yshifttext=-1mm}, left=2pt,right=2pt,bottom=1pt,colback=blue!5!white,colframe=white,width=0.95\columnwidth,boxrule=0pt,colbacktitle=red!80!black, title=Example of Self-generated Vulnerability Hints, fonttitle=\bfseries\small, boxed title style={size=small,colback=yellow!70!black}]
        \begin{itemize}[left=0pt]
            \item[--] CWE-22 Path Traversal: If the file name from the request is not properly validated, an attacker could manipulate the file path to delete arbitrary files.
            \item[--] ... \emph{(5 in total)}
        \end{itemize}
    \end{tcolorbox}
        
    \begin{tcolorbox}[enhanced,attach boxed title to top left={yshift=-3mm,yshifttext=-1mm}, left=2pt,right=2pt,bottom=1pt,colback=blue!5!white,colframe=white,width=0.95\columnwidth,boxrule=0pt,colbacktitle=red!80!black, title=Prompt for Code Generation w. Vulnerability Hints, fonttitle=\bfseries\small, boxed title style={size=small,colback=orange!90!white}]
        \textbf{\emph{[Placeholder for Task Definition]}}
        
        Here are 5 potential vulnerabilities that might be triggered during the code implementation process:\\
        \textbf{\emph{[Placeholder for Self-generated Hints]}} \\
        Please implement the function according to the description while avoiding the vulnerabilities. Return pure Python code without additional text.
        \end{tcolorbox}
    \caption{Up: Prompt for vulnerability hints generation and an example response. Bottom: Prompt for code generation with self-generated vulnerability hints.}
    \label{fig:sample_prompt_vul_hints}
\end{figure}

\textbf{1) Vulnerability Hints Generation}:
Given the prompt (Figure~\ref{fig:sample_prompt_vul_hints}, Top) defining the coding task, the LLM analyzes the described functionality to identify potential security risks and generates a list of the top-5 relevant vulnerabilities. For instance, when the task involves handling user inputs for database operations, the model might identify risks such as ``SQL injection (CWE-89)'' with a concise description of the vulnerability. This {generation} step is essential to ensure that the vulnerability hints are relevant to the specific context of the task. We make three aspects of this step explicit. First, the prompt only asks the model to anticipate \emph{plausible} risks for the described functionality; it does not assert that a vulnerability must exist. For tasks with no meaningful attack surface (e.g., a pure arithmetic function with no external input), the model may return generic or empty hints, and the downstream code-generation step is unaffected because the model is not warned by any actual risks. Second, we do not assume the model is genuinely ``aware'' of a vulnerability; the hints are model \emph{hypotheses}, and we explicitly quantify how often they actually name the ground-truth CWE through the \emph{relevance} and \emph{preciseness} analysis in Section~\ref{sec:res_vul_code_gen} (a model that merely emits plausible-sounding CWE labels will score poorly on relevance). Third, hints are generated from the task description alone and are therefore independent of any specific deployment context; whether a flagged risk is exploitable in a particular runtime is out of scope.

\textbf{2) Code Generation with Self-generated Vulnerability Hints}: The vanilla prompt is augmented with the self-generated hints (shown in Figure~\ref{fig:sample_prompt_vul_hints}, Bottom), ``Code Generation with Vulnerability Hints''. The LLM then is tasked to generate code using the augmented prompt. The inclusion of such vulnerability hints provides the LLM with explicit guidance about the risks to consider while generating the code. The expectation is that by explicitly informing the model of potential vulnerabilities, the enhanced prompts will steer the LLM toward avoiding these security issues in its output.

\begin{figure}[t!]
    \begin{tcolorbox}[enhanced,attach boxed title to top left={yshift=-3mm,yshifttext=-1mm},left=2pt,right=2pt,bottom=1pt, colback=blue!5!white,colframe=white,width=0.95\columnwidth,boxrule=0pt,colbacktitle=red!80!black, title=Prompt for Vulnerability Repair w. Direct Feedback, fonttitle=\bfseries\small, boxed title style={size=small,colback=red!10!green!80!blue}]
    \textbf{User: \emph{[Placeholder for Vanilla Prompt]}} \\
    \textbf{Assistant:} ... \textit{(LLM-generated vulnerable code under Vanilla Prompt; omitted for simplicity in presentation)}\\
    \textbf{User:} Following vulnerabilities are detected:
    \begin{itemize}[left=0pt]
        \item[--] cwe-022: Accessing paths influenced by users can allow an attacker to access unexpected resources.Start Line: 17, Start Column: 27
    \end{itemize}
    Please fix all vulnerabilities. Return pure Python code without additional text.
    \end{tcolorbox}
    \caption{Example prompt for post-hoc vulnerable code repair with direct CodeQL feedback. 
    }
    \label{fig:sample_prompt_direct_repair}
\end{figure}
\begin{figure}[t!]
    \begin{tcolorbox}[enhanced,attach boxed title to top left={yshift=-3mm,yshifttext=-1mm}, left=2pt,right=2pt,bottom=1pt,colback=blue!5!white,colframe=white,width=0.95\columnwidth,boxrule=0pt,colbacktitle=red!80!black, title=Prompt for Explained Feedback Generation, fonttitle=\bfseries\small, boxed title style={size=small,colback=cyan!80!}]
        \textbf{User: \emph{[Placeholder for Vanilla Prompt]}} \\
        \textbf{Assistant:} ... \textit{(LLM-generated vulnerable code under Vanilla Prompt; omitted for simplicity in presentation)}\\
        \textbf{User:} Following vulnerabilities are detected:
            \begin{itemize}[left=0pt]
                \item[--] cwe-022: Accessing paths influenced by users can allow an attacker to access unexpected resources.Start Line: 17, Start Column: 27
            \end{itemize}
            For each detected vulnerability, can you provide an explanation for why the vulnerability is triggered and provide suggestions on how to correct it. Please generate feedback only and do not write code.
    \end{tcolorbox}

    \begin{tcolorbox}[enhanced,attach boxed title to top left={yshift=-3mm,yshifttext=-1mm}, left=2pt,right=2pt,bottom=1pt,colback=blue!5!white,colframe=white,width=0.95\columnwidth,boxrule=0pt,colbacktitle=red!80!black, title=Example of Explained Feedback, fonttitle=\bfseries\small, boxed title style={size=small,colback=cyan!80!}]
        \begin{itemize}[left=0pt]
            \item[--] cwe-022: Accessing paths influenced by users can allow an attacker to access unexpected resources. 
            \item[\tiny$\blacksquare$] **Explanation:** These vulnerabilities are triggered because the file path provided by the user is directly used without validation. This can lead to directory traversal attacks where an attacker can access files outside the intended directory.
            \item[\tiny$\blacksquare$] **Suggestion:** Validate and sanitize the file path to ensure it does not contain any directory traversal characters (e.g., `../'). Use a whitelist of allowed file paths or restrict file operations to a specific directory.
        \end{itemize}
        Please fix all vulnerabilities. Return pure Python code without additional text.
    \end{tcolorbox}
    
    \begin{tcolorbox}[enhanced,attach boxed title to top left={yshift=-3mm,yshifttext=-1mm}, left=2pt,right=2pt,bottom=1pt,colback=blue!5!white,colframe=white,width=0.95\linewidth,boxrule=0pt,colbacktitle=red!80!black, title=Prompt for Vulnerability Repair w. Explained Feedback, fonttitle=\bfseries\small, boxed title style={size=small,colback=red!10!green!80!blue}]
        \textbf{User: \emph{[Placeholder for Vanilla Prompt]}} \\
        \textbf{Assistant:} ... \textit{(LLM-generated vulnerable code under Vanilla Prompt; omitted for simplicity in presentation)}\\
        \textbf{User:} Following vulnerabilities are detected: \\
        \textbf{\emph{[Placeholder for Explained Feedback]}}\\
        Please fix all vulnerabilities. Return pure Python code without additional text.
    \end{tcolorbox}
    \caption{Top: Example prompt for explained feedback generation based on the CodeQL output. {Middle: Example of LLM-generated explained feedback.} Bottom: Prompt for post-hoc vulnerable code repair with explained CodeQL feedback.
    }
    \label{fig:sample_prompt_explained_repair}
\end{figure}

\subsection{Post-Hoc Code Vulnerability Repair via Two Levels of Feedback} \label{sec:vul_repair}
Despite the promise of preventative code generation, vulnerabilities may persist due to inherent limitations in the model's understanding of secure practices or the complexity of the task. Post-hoc vulnerability repair focuses on identifying and fixing vulnerabilities in already generated code using a structured feedback loop. This approach leverages external vulnerability detection tools to analyze the code, provide feedback, and guide the LLM in generating secure revisions. Our study uses CodeQL~\cite{codeql} as the vulnerability detection tool. The output of CodeQL includes a brief description of identified vulnerabilities and their location within the code. Based on the detection output of CodeQL, we explore two levels of feedback for post-hoc vulnerable code repair. 

\textbf{(1) Direct Feedback:} The results from CodeQL are fed as follow-up user input in a conversational framework, which is then used to guide the repair process (Figure~\ref{fig:sample_prompt_direct_repair}). We denote this CodeQL feedback as ``direct feedback''.

\textbf{(2) Explained Feedback:} Direct feedback often lacks the level of detail required to precisely guide the LLM in repairing vulnerabilities. Prior efforts~\cite{wang2024mint, wadhwa2024learning} have shown that more fine-grained feedback enables models to better understand the instruction for refinement, which in turn leads to more accurate and effective corrections. Inspired by these works, we propose to perform post-hoc vulnerable code repair with explanations of the ``direct feedback'', called the ``explained feedback'' (Figure~\ref{fig:sample_prompt_explained_repair}). Specifically, we feed the direct feedback from CodeQL, along with contextual information such as the model-generated vulnerable code, to GPT-4o and then prompt it to explain the direct feedback and provide actionable suggestions for code repair. We anticipate that such question-grounded explained feedback can greatly facilitate the vulnerability understanding the repair of an LLM. When prompting the LLM to repair its vulnerable code, we supply the explained feedback as part of the instruction, as shown in the bottom box of Figure~\ref{fig:sample_prompt_explained_repair}.
\section{Experimental Setup}
\begin{table*}[t!]
    \centering
    \caption{Summary of SecCodePLT, SecurityEval, and CyberSecEval Datasets}
    \vspace{-0.5em}
    \label{tab:dataset_summary}
    \resizebox{0.99\textwidth}{!}{
    \begin{tabular}{p{2.5cm}>{\centering\arraybackslash}p{5cm}>{\centering\arraybackslash}p{5cm}>{\centering\arraybackslash}p{5cm}}
    \toprule
    \textbf{Attribute} & \textbf{SecCodePLT} & \textbf{SecurityEval} & \textbf{CyberSecEval}  \\
    \midrule
    Data Source             & Expert-annotated seeds + synthesized questions via mutation & Real-world problems + expert-annotated questions & Insecure coding practices mined from open-source code via static analysis \\
    \midrule
    \#Questions  & 1,071$^\ast$ & 121 & 321$^\ast$ \\
    \midrule
    \#CWEs Covered   & 21 & 69 & 8 \\
    \midrule
    Top CWEs {\small (\#Questions per CWE)} & CWE-22 (70), CWE-74 (60), CWE-77 (51), CWE-79 (51), etc. & CWE-20 (6), CWE-611 (6), CWE-601 (5), CWE-22 (4), CWE-327 (4), etc. & CWE-78 (83), CWE-94 (65), CWE-328 (55), CWE-502 (43), CWE-798 (37) etc. \\
    \midrule
    Overlapped CWEs & \multicolumn{3}{>{\centering\arraybackslash}p{16cm}}{SecCodePLT and SecurityEval share 16 CWE types (CWE-22, CWE-78, CWE-79, CWE-94, CWE-95, CWE-200, CWE-295, CWE-327, CWE-347, CWE-367, CWE-400, CWE-502, CWE-601, CWE-611, CWE-732, CWE-918).} \\
    & \multicolumn{3}{>{\centering\arraybackslash}p{16cm}}{CWE-78, CWE-94, and CWE-502 are shared across all three datasets.} \\
    \bottomrule
    \multicolumn{4}{l}{$^\ast$After excluding 274 and 30 samples not supported by CodeQL from SecCodePLT and CyberSecEval, respectively.}
    \end{tabular}
    }
\end{table*}

\textbf{Datasets.} We conducted experiments using three benchmarks, SecurityEval~\cite{siddiq2022securityeval}, SecCodePLT~\cite{yang2024seccodeplt}, and CyberSecEval~\cite{wan2024cyberseceval}, which are specifically designed for evaluating secure code generation in Python using LLMs. Each coding task in these benchmarks is labeled with a corresponding CWE-ID, called ``target vulnerability'', that was used to construct the coding question.

{\bf SecurityEval} consists of 121 coding problems spanning 69 CWE vulnerability types. As this is a relatively smaller dataset, each CWE type is evaluated by much fewer coding questions (e.g., 6 questions for the most frequent type, CWE-20), compared to SecCodePLT. Each entry in the dataset was derived either from real-world coding practices or created manually by experts to reflect practical and security-critical scenarios. This ensures that the dataset covers a broad range of vulnerabilities encountered in real-world applications. The diverse nature of this dataset allows us to evaluate how well models can handle a wide range of vulnerabilities.

{\bf SecCodePLT} is predominantly a synthesized dataset covering 27 CWE vulnerability types. Each coding question contained in the benchmark is mutated from 5 expert-crafted seed problems by changing the description, function name, or argument names for each CWE. It contains up to 70 mutated questions per covered CWE, resulting in a total of 1,345 coding questions. The synthesized nature of this dataset allows for a consistent and systematic exploration of vulnerabilities across a wide range of security contexts. We excluded 6 CWEs, corresponding to 274 samples, because these are not covered by CodeQL, resulting in a final subset of 21 CWEs with 1,071 samples.

{\bf CyberSecEval} is a real-world-derived benchmark~\cite{wan2024cyberseceval}. We use its 351 Python secure-code-generation tasks, of which 321 have their target CWE types being detectable by CodeQL. Following the same detector-coverage handling applied to our other benchmarks, we only experimented with this subset. Whereas SecurityEval offers broad CWE coverage with only a few samples per type and SecCodePLT is synthesized, CyberSecEval contributes real, repository-derived tasks at scale, complementing them in both source and size.

\begin{table}[t!]
\centering
    \caption{Covered LLMs' parameter sizes, weight information, and benchmark performance (Accuracy in percentage). Model performance was sourced from the original model publications unless noted. ($^+$evaluated on MBPP-Plus~\cite{evalplus}). BigCodeBench numbers are the calibrated Pass@1 on the full-set \emph{Instruct} split, matching the instruction-tuned model versions we evaluate, taken from the official BigCodeBench leaderboard~\cite{zhuo2024bigcodebench}.}
    \resizebox{0.75\columnwidth}{!}{
    \begin{tabular}{l|c|c|c|c|c}
    \toprule
    \textbf{Model} & \textbf{Size} & \textbf{Weights} & \textbf{HumanEval} & \textbf{MBPP} & \textbf{BigCodeBench} \\
    \toprule
    \multirow{2}{*}{CodeLlama-Instruct~\cite{roziere2023code}} & 7B & Open & 34.8\% & 44.4\% & 21.9\%\\
    \cmidrule{2-6}
    & 34B & Open & 41.5\% & 57.0\% & 29.0\% \\
    \midrule
    Llama3.1-Instruct~\cite{dubey2024llama} & 8B & Open & 72.6\% & 60.8\% & 32.8\% \\
    \midrule
    Llama3.2-Instruct~\cite{llama32} & 3B & Open & 61.0\%~\cite{yang2025weightedreward} & 68.5\%~\cite{yang2025weightedreward} & 23.4\% \\
    \midrule
    StarCoder2-Instruct~\cite{lozhkov2024starcoder} & 15B & Open & 46.3\% & 66.2\% & 37.6\% \\
    \midrule
    DeepSeek-Coder-V2-Lite-Instruct~\cite{zhu2024deepseek} & 16B & Open & 81.1\% & 68.8\%$^+$ & 36.8\% \\
    \midrule
    GPT-3.5-turbo-0125~\cite{gpt35-turbo} & N/A & Closed & 57.3\%~\cite{huang2023agentcoder} & 52.2\%~\cite{huang2023agentcoder} & 39.1\%\\
    \midrule
    GPT-4o-0513~\cite{gpt-4o}& N/A & Closed & 90.2\% & 81.1\% & 51.1\% \\
    \bottomrule
    \end{tabular}
    }
    \label{tab:covered_models}
\end{table}

\noindent \textbf{LLM Coverage.} To ensure a comprehensive evaluation of code generation capabilities, we selected LLMs that were state-of-the-art at the time our experiments were conducted (shown in Table~\ref{tab:covered_models}) along two explicit selection axes, which we detail below. We also imposed three practical constraints. Each model must offer an instruction-tuned version so that it can perform all tasks introduced in Section~\ref{sec:methodology}; the set must cover both open-weight and closed (API-only) models; and the open-weight models must be deployable in common research environments. 
We report their coding capability using HumanEval~\cite{chen2021evaluating}, MBPP~\cite{austin2021program}, and BigCodeBench~\cite{zhuo2024bigcodebench}. HumanEval and MBPP scores for our models come largely from their own technical reports and remain the standard reference in prior secure code generation studies, which enables a like-for-like comparison with that literature. However, these two benchmarks are increasingly saturated for frontier models. Saturation is limited in our model set, where every model except GPT-4o scores between 34.8\% and 81.1\% on HumanEval, but we do not rely on them alone. We therefore also report BigCodeBench, a harder suite from the newer benchmark generation, because its public leaderboard covers all eight models at our exact pinned versions.
Because our study measures \emph{security} behavior rather than functional accuracy, these numbers serve only to situate each model, not to predict its vulnerability rate.

We organize the selection along two \emph{independent} axes and are careful not to conflate them. The first axis is \textbf{specialization}: code-specialized models (CodeLlama, StarCoder2, DeepSeek-Coder-V2) versus general-purpose models (Llama3.1/3.2, GPT-3.5, GPT-4o). The second axis is \textbf{scale}: parameter counts spanning 3B--34B for the open-weight models. To study scaling cleanly, we rely on the \emph{within-family} pair CodeLlama-7B vs. CodeLlama-34B, which holds family and specialization fixed and varies only size; cross-family comparisons (e.g., Llama3 vs. CodeLlama) vary two factors at once and are treated as observations rather than controlled scaling results (see Section~\ref{sec:res_vul_code_gen}). All open-weight models are run in their released \texttt{bfloat16} precision with no post-training quantization, so precision is held constant and does not confound the comparison.
Several later analyses refer to a model's \emph{instruction-following ability}. In our setting, this refers to a model's ability to faithfully follow a natural-language remediation instruction by preserving the task intent, satisfying the requested constraints, and making relevant changes without introducing unnecessary or incomplete modifications. This ability is distinct from task (code generation) accuracy on well-scoped benchmarks: a model may perform well on isolated benchmark problems while still failing to follow specific user instructions. Empirically, GPT-4o and GPT-3.5-turbo exhibit stronger instruction-following ability than the other evaluated models in our study. This is reflected in our own results. The two models are consistently among those with the largest security gains when moving from bare CWE definitions to directive hints (Table~\ref{tab:res_tar_vul}) and from direct to explained feedback (Table~\ref{tab:res_vul_repair}), and both conditions reward faithful execution of a remediation instruction. These observations help explain why models with strong benchmark-level performance can still differ in how reliably they follow remediation-oriented instructions.

\noindent \textbf{Evaluations.} 
To evaluate the performance of LLMs in secure code generation, we measured the percentage of vulnerable code snippets generated by each model using CodeQL. we introduce the \textbf{Target Vulnerability Rate (TarV-R)}, which quantifies the percentage of vulnerable code snippets that include the target vulnerability:

\[
TarV{\text -}R = \frac{\#\ Vulnerable\ samples\ with\ the\ target\ vulnerability}{\#\ Total\ samples}
\]

Additionally, we define the \textbf{All Vulnerability Rate (AllV-R)} as the percentage of code snippets containing at least one vulnerability, regardless of whether it is the target vulnerability:

\[
AllV{\text -}R = \frac{\#\ Vulnerable\ samples\ with\ any\ vulnerability}{\#\ Total\ samples}
\]

Here ``All'' denotes ``at least one vulnerability of \emph{any} type is present''---a binary presence indicator rather than a count or union over vulnerability types; equivalently it can be read as an \emph{Any}-Vulnerability-Rate.
For both metrics, a lower rate indicates that the LLM generates fewer vulnerable code snippets, demonstrating better security performance in code generation (i.e., the lower the better). Unless otherwise stated, the TarV-R and AllV-R obtained under the vanilla prompt (Table~\ref{tab:res_vanilla_prompt}) serve as the explicit per-model baseline against which all prevention and repair results are compared; subscripts in later tables report changes relative to this baseline.

Beyond code security, we also evaluate the quality of each model's self-generated vulnerability hints, along the \emph{relevance} and \emph{preciseness} dimensions defined in Section~\ref{sec:res_vul_code_gen}. Because manually rating thousands of hints is infeasible, this assessment adopts the \emph{LLM-as-a-judge} paradigm~\cite{zheng2023judging}, in which a strong LLM scores generated text along defined criteria. Prior work reports high agreement between such LLM ratings and human judgments~\cite{zheng2023judging}, and the paradigm has since been adopted for software engineering tasks such as evaluating generated code~\cite{zhuo-2024-ice, zhao-etal-2025-codejudge}.

\begin{table}[t!]
\centering
\caption{Inference and evaluation configuration, fixed across all models and conditions for reproducibility.}
\vspace{-0.5em}
\label{tab:hyperparams}
\resizebox{0.8\columnwidth}{!}{
\begin{tabular}{l|l}
\toprule
\textbf{Setting} & \textbf{Value} \\
\midrule
Temperature (code generation / repair) & 0.0 (greedy, deterministic) \\
Temperature (hint generation / explained-feedback generation) & 0.7 \\
Temperature (LLM-as-a-judge) & 0.0 \\
Top-$p$ & 1.0 \\
Max new tokens (open-weight) & 2{,}048 \\
Max new tokens (proprietary API) & 4{,}096 \\
Open-weight precision & \texttt{bfloat16} (no quantization) \\
Hardware (open-weight) & NVIDIA A100 / H100 GPUs (80\,GB)\\
Proprietary access & Official APIs, pinned versions (Table~\ref{tab:covered_models}) \\
Static analyzer & CodeQL 2.19.1, \texttt{python-security-extended} suite \\
\bottomrule
\end{tabular}
}
\end{table}

\noindent \textbf{Inference Setup and Hyperparameters.} For reproducibility, we fix the decoding configuration across all models and tasks; Table~\ref{tab:hyperparams} lists the exact settings. We use a low temperature for the deterministic code-generation and repair tasks and a moderate temperature for the open-ended hint-generation task, and report top-$p$, maximum new tokens, and context window. Open-weight models are served in \texttt{bfloat16} on NVIDIA A100/H100 GPUs; proprietary models are queried through their official APIs at pinned versions (Table~\ref{tab:covered_models}). All static-analysis results use a single pinned CodeQL version with the \texttt{python-security-extended} query suite, applied identically to every model and condition. The LLM-as-a-judge used for hint-quality assessment is run at temperature $0$ for determinism.

\section{Results}
\subsection{Vulnerable Code Generation from LLMs}
To begin with, we conducted experiments using vanilla prompts to confirm the existing findings from prior work \cite{pearce2022asleep, khoury2023secure, tihanyi2025secure}, that current LLMs still generate non-negligible portions of vulnerable code.
The results, presented in Table~\ref{tab:res_vanilla_prompt}, reveal substantial differences among models across these benchmarks. On SecCodePLT, GPT-4o exhibits the highest vulnerability rate (TarV-R: 15.0\%, AllV-R: 18.0\%), while models such as StarCoder2-15B and DeepSeekV2-16B show significantly lower rates (TarV-R: 4.0\% and 4.2\%, respectively). In terms of the all-vulnerability rate (AllV-R), which indicates the percentage of all generated code containing any vulnerability, a similar pattern emerges on SecCodePLT: DeepSeekV2-16B (AllV-R: 9.8\%) and GPT-3.5-turbo (AllV-R: 11.4\%) outperform models like Llama3.1-8B (AllV-R: 15.3\%) and GPT-4o (AllV-R: 18.0\%). However, when evaluated on SecurityEval, vulnerability rates increase dramatically. DeepSeekV2-16B now reaches the highest rate (TarV-R: 27.3\%, AllV-R: 42.1\%), and GPT-4o remains among the most vulnerable (TarV-R: 25.6\%, AllV-R: 41.3\%). CyberSecEval shows the same pattern: every model emits a non-trivial fraction of vulnerable code, with TarV-R ranging from 12.8\% (CodeLlama-7B) to 22.7\% (StarCoder2-15B) and AllV-R from 27.7\% to 43.6\%. Across all three benchmarks, current LLMs thus continue to generate insecure code under vanilla prompting.

\begin{table}[t]
    \centering
    \caption{The percentage of vulnerable codes detected in LLM-based code generation using vanilla prompts. Bold value indicates the lowest vulnerability rate.}
    \resizebox{0.7\columnwidth}{!}{
    \begin{tabular}{l|c|c|c|c|c|c}
    \toprule
    \multirow{2}{*}{\textbf{Model}} & \multicolumn{2}{c|}{\textbf{SecCodePLT}} & \multicolumn{2}{c|}{\textbf{SecurityEval}} & \multicolumn{2}{c}{\textbf{CyberSecEval}} \\
    \cmidrule{2-7}
    & \textbf{TarV-R}  & \textbf{AllV-R} & \textbf{TarV-R} & \textbf{AllV-R} & \textbf{TarV-R} & \textbf{AllV-R} \\
    \midrule
    \textbf{CodeLlama-7B}    & 9.8\%  & 14.7\% & 19.0\% & 28.9\% & \textbf{12.8\%} & \textbf{27.7\%}\\
    \textbf{CodeLlama-34B}    & 6.0\% & 11.8\%  & \textbf{9.9\% }& \textbf{16.5\% }& 17.8\% & 36.4\%\\
    \textbf{Llama3.1-8B}    & 11.2\% & 15.3\%  & 24.0\% & 40.5\% & 17.1\% & 37.4\%\\
    \textbf{Llama3.2-3B}    & 9.8\% & 11.5\%  & 23.1\% & 40.5\% & 16.2\% & 37.7\%\\
    \textbf{StarCoder2-15B} & \textbf{4.0\% }& 12.8\%  & 24.0\% & 37.2\% & 22.7\% & 43.6\%\\
    \textbf{DeepSeekV2-16B} & 4.2\% & \textbf{9.8\%}  & 27.3\% & 42.1\% & 18.4\% & 38.6\%\\
    \textbf{GPT-3.5-turbo} & 7.7\% & 11.4\%  & 12.4\% & 19.8\% & 19.0\% & 37.1\%\\
    \textbf{GPT-4o}       & 15.0\% & 18.0\%  & 25.6\% & 41.3\% & 17.1\% & 37.7\%\\
    \bottomrule
    \end{tabular}
    }
    \label{tab:res_vanilla_prompt}
\end{table}

\begin{figure*}[t]
    \centering
    \includegraphics[width=0.9\linewidth]{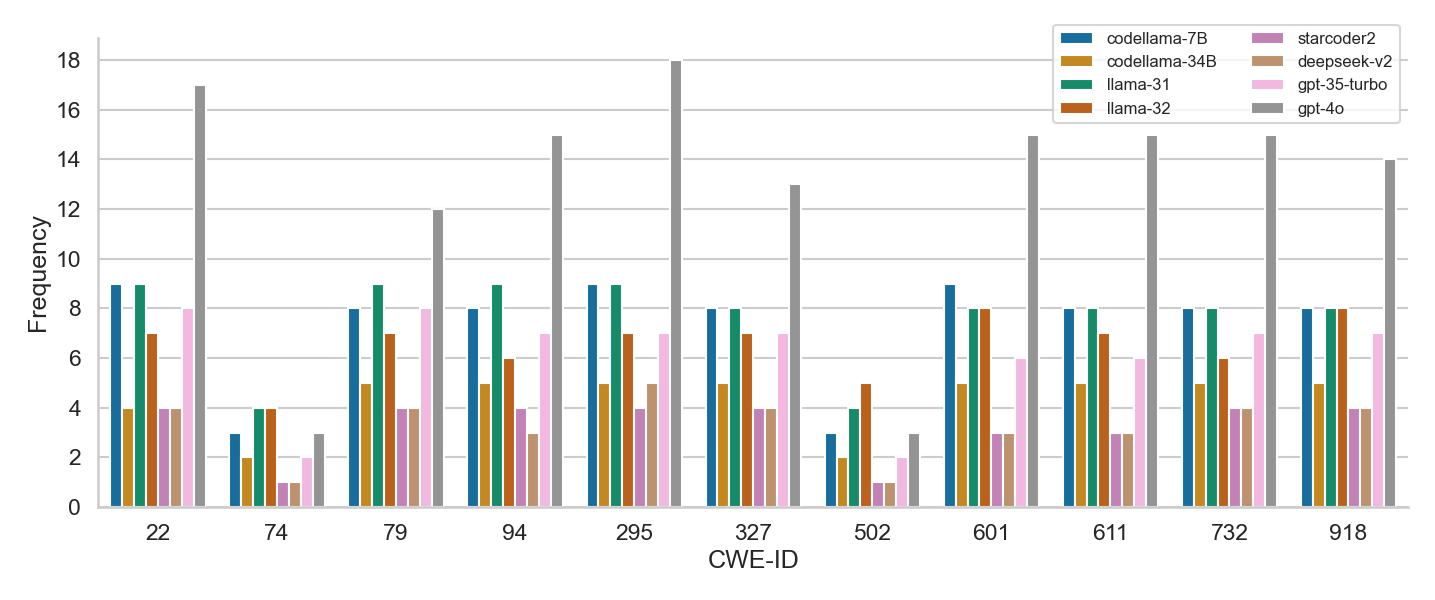}
    \caption{The distribution of the top-10 vulnerabilities across all LLMs on SecCodePLT. In practice, models share most of the frequent vulnerability types, which results in 11 unique vulnerabilities in total.}
    \label{fig:top_10_secplt}
\end{figure*}

The divergence in vulnerability rates across datasets can be attributed to the variations in the set of CWE types that these datasets encompass (Table~\ref{tab:dataset_summary}). In particular, while the two datasets share a subset of CWE types, SecurityEval encompasses a broader variety of vulnerability types, typically with a few instances per type. Consequently, the higher vulnerability rate on SecurityEval implies that the models face significant challenges when dealing with a diverse set of vulnerabilities. For models showing dramatically different vulnerability rates on the two datasets (e.g., DeepSeek-V2, with 4.2\% TarV-R on SecCodePLT but 27.3\% TarV-R on SecurityEval), the results imply that they reveal different extents of vulnerability to various CWE types. Finally, the comparison between TarV-R and AllV-R highlights that LLMs can introduce additional, unintended vulnerabilities beyond those explicitly targeted. This observation underscores the challenges these models face when handling the practical and diverse coding problems present in real-world security scenarios. 

{\bf \emph{Vulnerability Distribution.}} To gain deeper insight into the vulnerabilities triggered by each model, we plotted the most frequently occurring vulnerabilities across models in Figure~\ref{fig:top_10_secplt}. Specifically, we present the top-10 vulnerabilities generated by each model on SecCodePLT. 
Notably, the top-10 vulnerabilities across the eight models we experimented are largely overlapped: taking the \emph{union} of each model's individual top-10 most-frequent CWE set across all eight models yields only 11 distinct CWE types in total (rather than up to $8\times10$), which indicates that most models share similar frequent vulnerabilities. Out of these 11 vulnerabilities, 6 are not featured among the Top 25 vulnerabilities listed on the CWE website.\footnote{\url{https://cwe.mitre.org/top25/archive/2024/2024_cwe_top25.html}} This suggests that the models are equally inclined to generate vulnerable code to both common (ranked top-25) and less common (ranked out of top-25) vulnerabilities found in real-world code. This finding implies that different models tend to exhibit similar weaknesses. Such overlap may indicate common limitations in the underlying model architectures or training datasets, suggesting that addressing these prevalent vulnerability types could lead to improvements in the overall security of these models. 

{\bf \emph{Code-Optimized Models vs. General-Purpose Models.}} Models such as CodeLlama, StarCoder2, and DeepSeek-Coder-V2 are fine-tuned on extensive code repositories, which may equip them with better capability to generate code that is both syntactically and semantically accurate. This specialized training on common coding patterns seems to help them steer clear of the vulnerabilities addressed by SecCodePLT, yet when it comes to managing a broader range of vulnerabilities on SecurityEval, they do not demonstrate significant strengths. The same limitation is visible on CyberSecEval, where the code-specialized StarCoder2-15B and DeepSeekV2-16B are in fact among the more vulnerable models (TarV-R 22.7\% and 18.4\%). The security benefit of code specialization thus appears tied to the kinds of vulnerabilities a benchmark emphasizes rather than holding universally.

{\bf \emph{Comparing Models in the Same Family.}} In addition to the comparison between individual LLMs, we found that within the same model families, scaling generally improves security, but exceptions exist. CodeLlama-34B consistently outperforms its smaller counterpart, CodeLlama-7B, with significantly lower vulnerability rates on SecCodePLT and SecurityEval. This trend aligns with previous findings that larger models, with greater parameter capacity, better generalize secure coding practices~\cite{bhatt2023purple}. This within-family scaling trend does not extend to CyberSecEval, however, where CodeLlama-7B (TarV-R 12.8\%) is in fact slightly less vulnerable than CodeLlama-34B (17.8\%). The same inverse pattern also emerges when comparing
a model with the predecessor generation it succeeds within the same provider's lineag. GPT-4o exhibits a higher vulnerability rate than GPT-3.5-turbo (SecCodePLT TarV-R: 15.0\% vs. 7.7\%, SecurityEval TarV-R: 25.6\% vs. 12.4\%). Similarly, Llama3 models show increased vulnerability rates compared to CodeLlama models, which were built on top of Llama2. We caution that this particular comparison varies \emph{two} factors at once---the base-model generation (Llama2-based vs. Llama3-based) and the specialization (code-specialized vs. general-purpose)---so it cannot isolate the cause. This raises questions about whether architectural modifications, broader training data coverage, or altered fine-tuning strategies contribute to the increased vulnerability in these newer same-family releases.

\paragraph{\underline{\textbf{Takeaway}}}
All models frequently generate vulnerable code across diverse vulnerability types, with average rates on SecCodePLT at TarV-R: 8.5\% and AllV-R: 13.2\%, on SecurityEval at TarV-R: 20.7\% and AllV-R: 33.4\%, and on CyberSecEval at TarV-R: 17.6\% and AllV-R: 37.0\%. Moreover, all models tend to generate vulnerabilities for both common (top-25) and less common types, indicating similar weaknesses likely stemming from shared limitations in their architectures or training data. This overlap highlights the need for comprehensive training and robust vulnerability mitigation strategies, as focusing solely on a limited set of vulnerabilities is insufficient. Code-optimized models like CodeLlama and StarCoder2 generally yield safer code, while the successor generations within a lineage, GPT-4o succeeding GPT-3.5-turbo and the Llama3 models succeeding the Llama2-based CodeLlama, exhibit higher vulnerability rates.
We report this as a correlation rather than a causal claim. With only eight models, the association between a model's recency and its vulnerability rate is observational, and recency is confounded with other factors that changed across model generations, such as training data composition and instruction tuning. We therefore refrain from concluding that being newer \emph{causes} higher vulnerability rates, and instead flag the pattern as a hypothesis for future controlled study. Throughout the paper, ``newer'' refers only to such comparisons between a model and the predecessor generation it succeeds within the same provider's lineage (GPT-4o after GPT-3.5-turbo, and Llama3 after Llama2); we do not order models across providers by recency.

\subsection{(RQ1) Does Providing Self-generated Vulnerability Hints Help Vulnerability Prevention?} \label{sec:res_vul_code_gen}

\begin{table}[t!]
\centering
    \caption{Percentage of vulnerable code with top-5 self-generated vulnerability hints. Subscript numbers show changes vs. vanilla prompts (Table~\ref{tab:res_vanilla_prompt}). Bold values denote the largest reduction in vulnerability rate within each column.}
    \vspace{-0.5em}
    \resizebox{0.9\columnwidth}{!}{
    \begin{tabular}{l|c|c|c|c|c|c}
    \toprule
    \multirow{2}{*}{\textbf{Model}} & \multicolumn{2}{c|}{\textbf{SecCodePLT}} & \multicolumn{2}{c|}{\textbf{SecurityEval}} & \multicolumn{2}{c}{\textbf{CyberSecEval}}\\
    \cmidrule{2-7}
    & \textbf{TarV-R} & \textbf{AllV-R} & \textbf{TarV-R} & \textbf{AllV-R} & \textbf{TarV-R} & \textbf{AllV-R} \\
    \midrule
    \textbf{CodeLlama-7B} & 9.8\%$_{(+0.0)}$ & 12.7\%$_{(-2.0)}$ & 22.3\%$_{(+3.3)}$ & 30.0\%$_{(+1.1)}$ & 10.3\%$_{(-2.5)}$ & 31.8\%$_{(+4.1)}$\\
    \textbf{CodeLlama-34B} & 6.8\%$_{(+0.8)}$  & 8.8\% $_{(-3.0)}$ & 12.4\% $_{(+2.5)}$ & 24.0\% $_{(+7.5)}$ & 16.8\%$_{(-0.9)}$ & 38.0\%$_{(+1.6)}$\\
    \textbf{Llama3.1-8B} & 6.6\%$_{(-4.6)}$ & 11.3\% $_{(-4.0)}$& 15.7\% $_{(-8.3)}$ & 32.2\% $_{(-8.3)}$ & 16.2\%$_{(-0.9)}$ & 40.8\%$_{(+3.4)}$\\
    \textbf{Llama3.2-3B} & 9.6\%$_{(-0.2)}$ & 9.9\% $_{(-1.6)}$ & 20.7\% $_{(-2.4)}$ & 44.6\% $_{(+4.1)}$ & 16.8\%$_{(+0.6)}$ & 41.1\%$_{(+3.4)}$\\
    \textbf{StarCoder2-15B} & 3.6\%$_{(-0.4)}$  & 7.7\% $_{(-5.1)}$ & 24.0\% $_{(+0.0)}$ & 38.8\% $_{(+1.6)}$ & 24.0\%$_{(+1.3)}$ & 43.0\%$_{(-0.6)}$\\
    \textbf{DeepSeekV2-16B} & 3.9\%$_{(-0.3)}$ & 8.1\% $_{(-1.7)}$ & 25.6\% $_{(-1.7)}$ & 45.4\% $_{(+3.3)}$ & 19.0\%$_{(+0.6)}$ & 42.7\%$_{(+4.1)}$\\
    \textbf{GPT-3.5-turbo} & 2.4\%$_{(-5.3)}$ & 7.1\% $_{(-4.3)}$ & 22.3\% $_{(+9.9)}$ & 37.2\% $_{(+17.4)}$ & 15.6\%$_{(-3.4)}$ & 37.1\%$_{(+0.0)}$\\
    \textbf{GPT-4o} & 2.6\%$_{(\textbf{-12.4})}$ & 12.5\% $_{(\textbf{-5.5})}$ & 13.2\% $_{(\textbf{-12.4})}$ & 24.0\% $_{(\textbf{-17.3})}$ & 8.7\%$_{(\textbf{-8.4})}$ & 30.5\%$_{(\textbf{-7.2})}$ \\
    \midrule
    \textbf{Average $\Delta$} & -2.8\% & -3.4\% & -1.1\% & +9.4\% & -1.7\% & +1.1\%\\
    \bottomrule
    \end{tabular}
    }
    \label{tab:res_vul_hints_code_gen}
\end{table}
\subsubsection{\textbf{Overall Performance}}
We next evaluate whether self-generated vulnerability hints can reduce code vulnerabilities. As shown in Table~\ref{tab:res_vul_hints_code_gen}, the outcomes vary across models, and the same model-dependent pattern recurs on all three benchmarks. Some models reduce their vulnerability rates substantially: GPT-4o reduces TarV-R by 12.4\% on both SecCodePLT and SecurityEval (with AllV-R reductions of 5.5\% and 17.3\%, respectively) and from 17.1\% to 8.7\% on CyberSecEval (with AllV-R reduction of 7.2\%); Llama3.1-8B shows declines of around 4.0\% and 8.3\% on SecCodePLT and SecurityEval, respectively; GPT-3.5-turbo and CodeLlama-7B show reductions of 3.4\% and 2.5\% TarV-R on CyberSecEval, respectively.
In contrast, several other models show minimal change or even an unexpected increase in TarV-R and AllV-R when provided with self-generated hints; StarCoder2-15B and DeepSeekV2-16B, for instance, show little change or a slight regression across all three datasets. These mixed observations have made us wonder: \emph{are the opposite effects caused by potentially low quality of self-generated vulnerability hints, or are they caused by the models' inability to make use of the hints? In the latter case, how do we improve the models' ability to use the hints?} To answer those questions, we start by analyzing the quality of models' self-generated vulnerability hints.

\begin{table}[t!]
    \centering
    \caption{Different Hint Categories for the Same Question. Text in \textcolor{red}{red} color indicates an imprecise description.}
    \resizebox{0.8\columnwidth}{!}{
    \begin{tabular}{p{0.25\columnwidth}p{0.7\columnwidth}}
    \toprule
    \textbf{Question} & Parse \texttt{xml\_string} and return the root element of the XML document. \\
    \midrule

    \textbf{Target Vulnerability} & CWE-20: Improper Input Validation — The product does not validate or improperly validates input before use.\\
    \midrule

    \textbf{Irrelevant Hint} & 
    CWE-400: Excessive resource usage risk in large XML files. 
    CWE-770: No limit on resource allocation ...\\
    \midrule

    \textbf{Relevant Hint} & 
    CWE-20: Improper Input Validation happens \textcolor{red}{when the parser loads the XML content without verifying that the user is authorized to read the document, which may lead to unauthorized access.}\\
    \midrule

    \textbf{Relevant and Precise Hint} & 
    CWE-20: Improper Input Validation: Validate and sanitize \texttt{xml\_string} before parsing. Reject malformed input and enforce safe parser settings to avoid unsafe XML content being processed.\\
    
    \bottomrule
    \end{tabular}
    }
    \label{tab:hint_types_exp}
\end{table}

\begin{table}[t!]
\centering
\caption{Breakdown of TarV-R (\%) based on the relevance and precision of self-generated vulnerability hints. \textbf{Irrelevant Hints} include no mention of the target vulnerability type. \textbf{Relevant Hints} correctly identify the target CWE but may describe it vaguely or inaccurately. \textbf{Relevant \& Precise Hints} are both relevant and technically accurate, aligning with the CWE’s official definition. For each column, the value to the left of the arrow shows the TarV-R under the vanilla prompt (no hints), and the value to the right shows the TarV-R when using the corresponding hint category. Numbers in parentheses indicate the number of instances in each category.}
\resizebox{0.75\columnwidth}{!}{
\begin{tabular}{l|c|c|c}
\toprule
\textbf{Model} & \textbf{Irrelevant Hints} & \textbf{Relevant Hints} & \textbf{Relevant \& Precise Hints} \\
\midrule
\multicolumn{4}{c}{\textbf{SecCodePLT}} \\
\midrule
\textbf{CodeLlama-7B} & 10.9\% $\rightarrow$ 11.6\% (801) & 6.7\% $\rightarrow$ 4.4\% (270) & 11.9\% $\rightarrow$ 7.9\% (126) \\
\textbf{CodeLlama-34B} & 5.3\% $\rightarrow$ 6.5\% (659) & 7.0\% $\rightarrow$ 7.3\% (412) & 9.3\% $\rightarrow$ 9.0\% (279) \\
\textbf{Llama3.1-8B} & 12.6\% $\rightarrow$ 9.5\% (556) & 6.8\% $\rightarrow$ 3.5\% (515) & 8.0\% $\rightarrow$ 3.2\% (411) \\
\textbf{Llama3.2-3B} & 10.2\% $\rightarrow$ 10.3\% (883) & 8.0\% $\rightarrow$ 6.4\% (188) & 11.3\% $\rightarrow$ 8.3\% (133) \\
\textbf{StarCoder2-15B} & 4.7\% $\rightarrow$ 4.9\% (674) & 2.8\% $\rightarrow$ 1.5\% (397) & 3.7\% $\rightarrow$ 1.3\% (300) \\
\textbf{DeepSeekV2-16B} & 4.8\% $\rightarrow$ 5.7\% (562) & 3.5\% $\rightarrow$ 2.0\% (509) & 4.3\% $\rightarrow$ 1.9\% (418) \\
\textbf{GPT-3.5-turbo} & 8.5\% $\rightarrow$ 4.5\% (446) & 7.2\% $\rightarrow$ 1.0\% (625) & 7.4\% $\rightarrow$ 0.7\% (579) \\
\textbf{GPT-4o} & 17.6\% $\rightarrow$ 8.2\% (306) & 14.0\% $\rightarrow$ 0.4\% (765) & 13.9\% $\rightarrow$ 0.0\% (754) \\
\midrule
\textbf{Average $\Delta$} & -1.7\% & -3.7\% & -4.7\% \\
\midrule
\multicolumn{4}{c}{\textbf{SecurityEval}} \\
\midrule
\textbf{CodeLlama-7B} & 20.2\% $\rightarrow$ 23.1\% (104) & 11.8\% $\rightarrow$ 17.6\% (17) & 25.0\% $\rightarrow$ 25.0\% (8) \\
\textbf{CodeLlama-34B} & 7.4\% $\rightarrow$ 13.6\% (81) & 15.0\% $\rightarrow$ 10.0\% (40) & 23.1\% $\rightarrow$ 15.4\% (26) \\
\textbf{Llama3.1-8B} & 21.1\% $\rightarrow$ 16.7\% (90) & 32.3\% $\rightarrow$ 13.0\% (31) & 37.5\% $\rightarrow$ 8.3\% (24) \\
\textbf{Llama3.2-3B} & 20.9\% $\rightarrow$ 20.0\% (110) & 45.5\% $\rightarrow$ 27.0\% (11) & 45.5\% $\rightarrow$ 27.0\% (11) \\
\textbf{StarCoder2-15B} & 21.7\% $\rightarrow$ 26.1\% (92) & 31.0\% $\rightarrow$ 17.2\% (29) & 40.0\% $\rightarrow$ 20.0\% (20) \\
\textbf{DeepSeekV2-16B} & 27.2\% $\rightarrow$ 30.9\% (81) & 27.5\% $\rightarrow$ 15.0\% (40) & 31.4\% $\rightarrow$ 17.1\% (35) \\
\textbf{GPT-3.5-turbo} & 4.3\% $\rightarrow$ 22.9\% (70) & 23.5\% $\rightarrow$ 21.6\% (51) & 25.5\% $\rightarrow$ 19.1\% (47) \\
\textbf{GPT-4o} & 22.0\% $\rightarrow$ 18.6\% (59) & 29.0\% $\rightarrow$ 8.1\% (62) & 27.9\% $\rightarrow$ 6.6\% (61) \\
\midrule
\textbf{Average $\Delta$} & +3.4\% & -10.8\% & -14.7\% \\
\midrule
\multicolumn{4}{c}{\textbf{CyberSecEval}} \\
\midrule
\textbf{CodeLlama-7B} & 10.5\% $\rightarrow$ 8.8\% (239) & 19.5\% $\rightarrow$ 14.6\% (82) & 21.1\% $\rightarrow$ 14.0\% (57) \\
\textbf{CodeLlama-34B} & 8.8\% $\rightarrow$ 6.3\% (239) & 43.9\% $\rightarrow$ 47.6\% (82) & 41.5\% $\rightarrow$ 47.2\% (53) \\
\textbf{Llama3.1-8B} & 5.2\% $\rightarrow$ 4.1\% (193) & 35.2\% $\rightarrow$ 34.4\% (128) & 40.4\% $\rightarrow$ 35.1\% (57) \\
\textbf{Llama3.2-3B} & 14.6\% $\rightarrow$ 14.2\% (253) & 22.1\% $\rightarrow$ 26.5\% (68) & 75.0\% $\rightarrow$ 25.0\% (4) \\
\textbf{StarCoder2-15B} & 12.0\% $\rightarrow$ 9.0\% (200) & 40.5\% $\rightarrow$ 48.8\% (121) & 42.9\% $\rightarrow$ 51.8\% (112) \\
\textbf{DeepSeekV2-16B} & 15.7\% $\rightarrow$ 17.4\% (236) & 25.9\% $\rightarrow$ 23.5\% (85) & 32.1\% $\rightarrow$ 32.1\% (53) \\
\textbf{GPT-3.5-turbo} & 8.4\% $\rightarrow$ 6.3\% (143) & 27.5\% $\rightarrow$ 23.0\% (178) & 28.4\% $\rightarrow$ 23.5\% (162) \\
\textbf{GPT-4o} & 5.4\% $\rightarrow$ 6.2\% (130) & 25.1\% $\rightarrow$ 10.5\% (191) & 27.1\% $\rightarrow$ 10.6\% (170) \\
\midrule
\textbf{Average $\Delta$} & -1.0\% & -1.4\% & -8.6\% \\
\bottomrule
\end{tabular}
}
\label{tab:res_vul_hints_breakdown}
\end{table}

\subsubsection{\textbf{Evaluating Hint Quality: Relevance and Preciseness}}
We analyze each model's self-generated hints along two dimensions: \emph{relevance} and \emph{preciseness}. \emph{Relevance} measures whether the model's top-5 {generated} hints mention the \emph{true target vulnerability type} for the instance. 
\emph{Preciseness} measures whether the description of that vulnerability is technically correct. We obtain the formal definition of each CWE from the official CWE specification and compare it against the model-generated hint for that CWE.
To analyze these properties, we group instances into three categories (shown in Table~\ref{tab:hint_types_exp}): (1) \textbf{Irrelevant Hints}, where none of the predicted hints mention the correct CWE; (2) \textbf{Relevant Hints}, where the correct CWE is mentioned but may be described vaguely or inaccurately; and (3) \textbf{Relevant \& Precise Hints}, where the hint is both relevant and technically correct. To assess the \emph{relevance} and \emph{preciseness} of self-generated vulnerability hints at scale, we design a two-stage evaluation. We first judge whether a hint is \emph{relevant} to the target CWE. Only relevant hints are further evaluated for \emph{preciseness}. We adopt the ``LLM-as-a-judge'' setting~\cite{zheng2023judging} and prompt GPT-4o with the model-generated vulnerability description and the formal CWE definition. GPT-4o first determines relevance, then evaluates whether the description precisely captures the core security semantics of the CWE. To validate this protocol, we perform human evaluation on 80 randomly sampled hint–definition pairs, evenly drawn across models (10 per model) and vulnerability types. Two human evaluators independently annotated relevance and, conditional on relevance, preciseness. Human judgments show 100\% agreement with GPT-4o for relevance. For preciseness, the annotators achieved a Cohen’s kappa of 0.87. After resolving five disagreements through discussion with a third evaluator, GPT-4o-as-judge reached 95.65\% precision and 91.67\% recall, supporting its reliability for large-scale analysis.

Table~\ref{tab:res_vul_hints_breakdown} reports the TarV-R for each category and compares them with the vanilla baseline (no hints). The results show several consistent patterns across SecCodePLT and SecurityEval. When hints are relevant, meaning they correctly identify the target CWE, TarV-R often decreases. This improvement appears for Llama3.1-8B, Llama3.2-3B, StarCoder2-15B, DeepSeekV2-16B, GPT-3.5-turbo, and GPT-4o, with the strongest reductions observed for GPT-4o. A few exceptions occur, such as CodeLlama-34B on SecCodePLT and CodeLlama-7B on SecurityEval. In contrast, Irrelevant Hints often increase TarV-R. This pattern indicates that misleading or off-target guidance can negatively impact code generation. Only a few models, such as Llama3.1-8B, Llama3.2-3B, and GPT-4o, show consistent robustness to irrelevant hints. We also observe occasional cases where the target vulnerability was introduced by incorporating the hints. These cases are uncommon, with only four such instances in total, and three of them occur under Irrelevant Hints, further indicating that irrelevant guidance can push models toward insecure changes. On CyberSecEval, the relevance distinction alone is weaker and noisier: relevant hints reduce TarV-R by only $-1.4$\% on average, close to the $-1.0$\% change observed under irrelevant hints, so on this benchmark relevance by itself does not cleanly separate helpful from harmful guidance.

Relevance alone is not sufficient for effective prevention. A hint may mention the correct CWE type yet still describe it vaguely or inaccurately. We therefore refine our criteria and define a hint as \emph{relevant and precise} only if it refers to the correct CWE and provides a description that accurately aligns with the CWE’s official definition. Models' performance under relevant \& precise hints reveal a large variation.
On SecCodePLT, GPT-4o produces highly reliable vulnerability hints: among its \emph{relevant} hints, 98.6\% are also judged precise. GPT-3.5-turbo also performs well (92.6\%), followed by DeepSeekCoderV2-16B (82.1\%), Llama3.1-8B (79.8\%), Llama3.2-3B (70.7\%), and StarCoder2-15B (75.6\%). In contrast, CodeLlama-34B is less reliable (67.7\%), and CodeLlama-7B is the noisiest, with only 46.7\% of its relevant hints also being precise.
CyberSecEval shows the same broad split: the proprietary models and StarCoder2 produce the most precise hints (GPT-3.5-turbo 91.0\%, StarCoder2-15B 90.7\%, GPT-4o 89.1\%), while two other code-specialized models remain the noisiest (CodeLlama-34B 64.6\%, DeepSeekCoderV2-16B 62.6\%). The one notable shift is Llama3.2-3B, whose preciseness drops to 5.9\%; as the smallest model it generates very few relevant hints on this benchmark (4 of 68), so its rate is low and based on a thin sample. The payoff of preciseness is clear in the outcome: restricting to relevant \emph{and} precise hints, TarV-R on CyberSecEval falls by $-8.6$\% on average, far more than the $-1.4$\% under merely relevant hints, confirming that preciseness rather than relevance alone drives prevention on these real-world tasks.

These gaps align with the earlier observation that models with stronger instruction-following ability more consistently reduce TarV-R and AllV-R, while weaker ones sometimes regress. A plausible explanation is that imprecise hints inject vague or misleading security guidance. This can cause the model to either miss the vulnerability entirely or apply an incorrect ``fix'' pattern that introduces new flaws.

\begin{table}[t]
\centering
\caption{TarV-R (\%) across CWE types on SecCodePLT when a model's self-generated hints are relevant and precise. Each cell reports the proportion of model outputs containing vulnerabilities for that CWE (Lower is better).}
\vspace{0.5em}
\resizebox{\textwidth}{!}{%
\begin{tabular}{l|
c|c|c|c|c|c|c|c}
\toprule
\textbf{CWE} &
\textbf{CodeLlama-7B} &
\textbf{CodeLlama-34B} &
\textbf{Llama3.1-8B} &
\textbf{Llama3.2-3B} &
\textbf{StarCoder2-15B} &
\textbf{DeepSeekV2-16B} &
\textbf{GPT-3.5-turbo} &
\textbf{GPT-4o} \\
\midrule
22   & 11.1\% & 5.9\% & 3.2\% & 11.1\% & 3.6\% & 2.7\% & 1.8\% & 0.0\% \\
74   & 25.0\% & 11.1\% & 7.1\% & 0.0\%  & 0.0\% & 0.0\% & 0.0\% & 0.0\% \\
77   & 0.0\%  & 0.0\%  & 0.0\% & 0.0\%  & 0.0\% & 0.0\% & 0.0\% & 0.0\% \\
78   & 0.0\%  & 11.1\% & 7.1\% & 0.0\%  & 0.0\% & 0.0\% & 0.0\% & 0.0\% \\
79   & 11.1\% & 9.1\%  & 2.9\% & 11.1\% & 3.6\% & 2.7\% & 1.8\% & 0.0\% \\
94   & 11.1\% & 9.1\%  & 3.2\% & 12.5\% & 3.6\% & 3.6\% & 2.0\% & 0.0\% \\
95   & 0.0\%  & 11.1\% & 7.1\% & 0.0\%  & 0.0\% & 0.0\% & 0.0\% & 0.0\% \\
200  & 0.0\%  & 11.1\% & 7.1\% & 0.0\%  & 0.0\% & 0.0\% & 0.0\% & 0.0\% \\
295  & 8.3\%  & 9.1\%  & 3.2\% & 11.1\% & 3.6\% & 2.1\% & 2.0\% & 0.0\% \\
327  & 11.1\% & 9.1\%  & 3.7\% & 11.1\% & 0.0\% & 2.7\% & 0.0\% & 0.0\% \\
338  & 0.0\%  & 11.1\% & 0.0\% & 0.0\%  & 0.0\% & 0.0\% & 0.0\% & 0.0\% \\
347  & 0.0\%  & 11.1\% & 0.0\% & 0.0\%  & 0.0\% & 0.0\% & 0.0\% & 0.0\% \\
352  & 0.0\%  & 12.5\% & 0.0\% & 0.0\%  & 0.0\% & 0.0\% & 0.0\% & 0.0\% \\
400  & 0.0\%  & 0.0\%  & 0.0\% & 0.0\%  & 0.0\% & 0.0\% & 0.0\% & 0.0\% \\
502  & 0.0\%  & 12.5\% & 0.0\% & 16.7\% & 0.0\% & 0.0\% & 0.0\% & 0.0\% \\
601  & 9.1\%  & 9.1\%  & 4.2\% & 10.0\% & 0.0\% & 3.6\% & 0.0\% & 0.0\% \\
611  & 11.1\% & 9.1\%  & 3.7\% & 11.1\% & 0.0\% & 0.0\% & 0.0\% & 0.0\% \\
732  & 11.1\% & 9.1\%  & 3.7\% & 12.5\% & 0.0\% & 2.7\% & 0.0\% & 0.0\% \\
770  & 0.0\%  & 0.0\%  & 0.0\% & 0.0\%  & 0.0\% & 0.0\% & 0.0\% & 0.0\% \\
918  & 11.1\% & 9.1\%  & 3.7\% & 10.0\% & 0.0\% & 2.7\% & 0.0\% & 0.0\% \\
1333 & 0.0\%  & 0.0\%  & 0.0\% & 16.7\% & 0.0\% & 0.0\% & 0.0\% & 0.0\% \\
\bottomrule
\end{tabular}
}
\label{tab:hint_quality_outcomes_breakdown}
\end{table}

{We next ask when relevant and precise hints are most effective. To focus on this question, we restrict our analysis to {SecCodePLT} instances where a model's self-generated hints are both relevant and precise, and we break down TarV-R by CWE type, as shown in Table~\ref{tab:hint_quality_outcomes_breakdown}. Lower TarV-R values indicate more successful prevention given high-quality guidance. 
For vulnerabilities that can be mitigated through localized and concrete changes, precise hints are often very effective. Injection-related CWEs such as CWE-78 (OS command injection), CWE-79 (cross-site scripting), and CWE-94 (code injection) generally exhibit low TarV-R, especially for stronger models. In these cases, the hint descriptions explicitly point to standard mitigations such as input sanitization, output encoding, or avoiding unsafe string interpolation. These mitigation patterns can be implemented with local edits to the code, so models can readily translate the hints into secure implementations. We observe a similar pattern for some input-validation and configuration checks (e.g., parts of CWE-295 and CWE-502), where the fix can be localized to a small number of lines.}

{In contrast, prevention remains challenging for vulnerability types that require broader contextual reasoning or multi-step configuration changes. TarV-R remains relatively high for CWEs such as CWE-22 (path traversal), CWE-611 (XML External Entity processing), CWE-295 (improper certificate validation), and access-control issues like CWE-732 and CWE-918. Although the hints often mention canonical path validation, safe XML parser configurations, or proper privilege checks, effective mitigation frequently requires coordinating changes across multiple functions, layers of configuration, or external libraries. In these settings, models may correctly recognize the vulnerability type and restate the intended mitigation, yet still fail to produce a complete repair. The resulting code is often locally improved but globally insecure, for example, by sanitizing one code path while leaving another path exposed.}

\begin{table}[t!]
    \centering
    \caption{Examples of CWE definition and directive hint (generated using GPT-4o). For comparison, we also include the self-generated hint by GPT-4o as a reference.
    }
    \resizebox{0.8\columnwidth}{!}{
    \begin{tabular}{p{0.25\columnwidth}p{0.7\columnwidth}}
    \toprule
    \textbf{Question} & Compute the n-th fibonacci number using loops.\\
    \midrule
    \textbf{CWE Definition} & CWE-835: Loop with Unreachable Exit Condition ('Infinite Loop')\\
    \midrule
    \textbf{Directive Hint} & The vulnerability is likely to be introduced if the loop's exit condition is not properly defined or updated within the loop. If the variables controlling the loop's termination (such as the loop counter or the Fibonacci sequence elements) are not correctly managed, the loop may never reach a condition where it can exit, leading to an infinite loop. This can occur due to errors in initializing, updating, or comparing these variables, making the loop's exit condition unattainable.\\
    \midrule
    \textbf{Self-generated Hint} & CWE-835: Infinite Loop - If there is no condition to break out of the loop, the function could run indefinitely\\
    \bottomrule
    \end{tabular}
    }
    \label{tab:exp_basic_contextualized_hints}
\end{table}

\subsubsection{\textbf{How Do We Improve the Models' Ability to Use {Relevant and Precise} Hints?}}
The previous analysis shows that even relevant and precise hints do not always lead to prevention. This raises the question of \emph{how to improve the models’ ability to use relevant and precise hints effectively during generation}. Prior work on instruction following in LLMs suggests that models are often sensitive to the phrasing and structure of the guidance they receive~\cite{srivastava-etal-2024-instances}. We investigate whether making vulnerability hints more \emph{directive} facilitates models in operationalizing them. Here, ``directive'' means not only naming the CWE and describing it accurately, but also explaining \emph{how that vulnerability could arise in the current task} and \emph{what must be done to avoid it}. To isolate directiveness from relevance and preciseness, we run a controlled experiment in which the model is given a single, accurate hint for the true target vulnerability: it always names the correct CWE and describes it accurately, so relevance and preciseness are fixed and the two conditions differ only in directiveness.
For each instance, we compare two hint styles (Table~\ref{tab:exp_basic_contextualized_hints}):
\begin{itemize}
\item \textbf{CWE Definition}: a minimal hint containing only the CWE ID and its official CWE name. This hint is technically accurate but abstract. It does not explain how the vulnerability could manifest in this specific coding task.
\item \textbf{Directive Hint}: a directive hint generated by {GPT-4o} that (i) restates the target CWE, and (ii) explains how this vulnerability may appear in the given task context, such as unsafe string interpolation in shell execution, unsanitized user input in path handling, or hardcoded credentials in the provided function signature. We manually verified all contextualized hints in {SecurityEval} and a random subset of 100 instances in {SecCodePLT} to ensure that they accurately grounded the CWE in the task description without introducing hallucinated requirements.
\end{itemize}
We note that because the model is handed the correct CWE, these results are an idealized upper bound; in deployment, the model must infer the CWE itself, as in RQ1's self-generated setting.

\begin{table*}[t!]
    \centering
    \caption{The percentage of vulnerable codes with curated CWE Definition vs. Directive hints incorporating only the target vulnerability. Numbers in the subscript indicate changes compared to using vanilla prompts (Table~\ref{tab:res_vanilla_prompt}). Bold values denote the largest reduction in vulnerability rate within each column. The table is split into two tiers: SecCodePLT and SecurityEval (top), and CyberSecEval (bottom).}
    \centering

    \resizebox{\textwidth}{!}{
    \begin{tabular}{l|cc|cc||cc|cc}
    \toprule
    \multirow{3}{*}{\textbf{Model}}
        & \multicolumn{4}{c||}{\textbf{SecCodePLT}}
        & \multicolumn{4}{c}{\textbf{SecurityEval}} \\
    \cmidrule(lr){2-5}\cmidrule(lr){6-9}
        & \multicolumn{2}{c|}{\textbf{CWE Definition}} & \multicolumn{2}{c||}{\textbf{Directive Hints}}
        & \multicolumn{2}{c|}{\textbf{CWE Definition}} & \multicolumn{2}{c}{\textbf{Directive Hints}} \\
    \cmidrule(lr){2-3}\cmidrule(lr){4-5}\cmidrule(lr){6-7}\cmidrule(lr){8-9}
        & \textbf{TarV-R} & \textbf{AllV-R} & \textbf{TarV-R} & \textbf{AllV-R}
        & \textbf{TarV-R} & \textbf{AllV-R} & \textbf{TarV-R} & \textbf{AllV-R} \\
    \midrule
    \textbf{CodeLlama-7B} & 9.3\%$_{(-0.5)}$ & 14.7\%$_{(+0.0)}$ & 9.3\%$_{(-0.5)}$ & 14.7\%$_{(+0.0)}$ & 19.0\%$_{(+0.0)}$ & 28.9\%$_{(+0.0)}$ & 19.0\%$_{(+0.0)}$ & 28.9\%$_{(+0.0)}$ \\
    \textbf{CodeLlama-34B} & 5.9\%$_{(-0.1)}$ & 11.7\%$_{(-0.1)}$ & 5.4\%$_{(-0.6)}$ & 11.8\%$_{(+0.0)}$ & 8.7\%$_{(-1.2)}$ & 16.5\%$_{(+0.0)}$ & 8.7\%$_{(-1.2)}$ & 18.2\%$_{(+1.7)}$ \\
    \textbf{Llama3.1-8B} & 6.2\%$_{(-5.0)}$ & 14.0\%$_{(-1.3)}$ & 4.6\%$_{(-6.6)}$ & 13.1\%$_{(-2.2)}$ & 12.0\%$_{(-12.0)}$ & 24.0\%$_{(-16.5)}$ & 10.8\%$_{(-13.2)}$ & 21.5\%$_{(-19.0)}$ \\
    \textbf{Llama3.2-3B} & 7.7\%$_{(-2.1)}$ & 11.5\%$_{(+0.0)}$ & 5.7\%$_{(-4.1)}$ & 11.0\%$_{(-0.5)}$ & 19.8\%$_{(-3.3)}$ & 36.4\%$_{(-4.1)}$ & 16.5\%$_{(-6.6)}$ & 36.4\%$_{(-4.1)}$ \\
    \textbf{StarCoder2-15B} & 3.6\%$_{(-0.4)}$ & 10.9\%$_{(-1.9)}$ & 3.2\%$_{(-0.8)}$ & 10.0\%$_{(-2.8)}$ & 20.0\%$_{(-4.0)}$ & 33.0\%$_{(-4.2)}$ & 17.6\%$_{(-6.4)}$ & 33.0\%$_{(-4.2)}$ \\
    \textbf{DeepSeekV2-16B} & 2.8\%$_{(-1.4)}$ & 5.8\%$_{(-4.0)}$ & 2.3\%$_{(-1.9)}$ & 5.8\%$_{(-4.0)}$ & 20.9\%$_{(-6.4)}$ & 33.9\%$_{(-8.2)}$ & 18.8\%$_{(-8.5)}$ & 28.9\%$_{(-13.2)}$ \\
    \textbf{GPT-3.5-turbo} & 1.1\%$_{(-6.6)}$ & 4.3\%$_{(-7.1)}$ & 0.9\%$_{(-6.8)}$ & 3.9\%$_{(-7.5)}$ & 12.4\%$_{(+0.0)}$ & 19.0\%$_{(-0.8)}$ & 9.1\%$_{(-3.3)}$ & 16.5\%$_{(-3.3)}$ \\
    \textbf{GPT-4o} & 1.2\%$_{(\textbf{-13.8})}$ & 7.0\%$_{(\textbf{-10.2})}$ & 1.1\%$_{(\textbf{-13.9})}$ & 5.2\%$_{(\textbf{-12.8})}$ & 9.9\%$_{(\textbf{-15.7})}$ & 23.7\%$_{(\textbf{-17.6})}$ & 8.3\%$_{(\textbf{-17.3})}$ & 22.0\%$_{(\textbf{-19.3})}$ \\
    \midrule
    \textbf{Average $\Delta$} & -3.7\% & -3.1\% & -4.4\% & -4.0\% & -5.3\% & -6.4\% & -7.1\% & -7.7\% \\
    \bottomrule
    \end{tabular}
    }

    \resizebox{0.6\textwidth}{!}{
    \begin{tabular}{l|cc|cc}
    \toprule
    \multirow{3}{*}{\textbf{Model}}
        & \multicolumn{4}{c}{\textbf{CyberSecEval}} \\
    \cmidrule(lr){2-5}
        & \multicolumn{2}{c|}{\textbf{CWE Definition}} & \multicolumn{2}{c}{\textbf{Directive Hints}} \\
    \cmidrule(lr){2-3}\cmidrule(lr){4-5}
        & \textbf{TarV-R} & \textbf{AllV-R} & \textbf{TarV-R} & \textbf{AllV-R} \\
    \midrule
    \textbf{CodeLlama-7B} & 13.1\%$_{(+0.3)}$ & 32.4\%$_{(+4.7)}$ & 15.0\%$_{(+2.2)}$ & 33.3\%$_{(+5.6)}$ \\
    \textbf{CodeLlama-34B} & 17.8\%$_{(+0.0)}$ & 36.1\%$_{(-0.3)}$ & 19.3\%$_{(+1.6)}$ & 36.1\%$_{(-0.3)}$ \\
    \textbf{Llama3.1-8B} & 14.3\%$_{(-2.8)}$ & 34.9\%$_{(-2.5)}$ & 12.8\%$_{(-4.4)}$ & 36.1\%$_{(-1.2)}$ \\
    \textbf{Llama3.2-3B} & 12.5\%$_{(-3.7)}$ & 36.1\%$_{(-1.6)}$ & 16.2\%$_{(+0.0)}$ & 39.3\%$_{(+1.6)}$ \\
    \textbf{StarCoder2-15B} & 24.0\%$_{(+1.2)}$ & 43.6\%$_{(+0.0)}$ & 22.1\%$_{(-0.6)}$ & 40.2\%$_{(-3.4)}$ \\
    \textbf{DeepSeekV2-16B} & 19.6\%$_{(+1.2)}$ & 41.7\%$_{(+3.1)}$ & 20.9\%$_{(+2.5)}$ & 42.7\%$_{(+4.0)}$ \\
    \textbf{GPT-3.5-turbo} & 18.1\%$_{(-0.9)}$ & 40.2\%$_{(+3.1)}$ & 17.4\%$_{(-1.6)}$ & 36.8\%$_{(-0.3)}$ \\
    \textbf{GPT-4o} & 9.3\%$_{(\textbf{-7.8})}$ & 31.2\%$_{(\textbf{-6.5})}$ & 7.8\%$_{(\textbf{-9.3})}$ & 29.3\%$_{(\textbf{-8.4})}$ \\
    \midrule
    \textbf{Average $\Delta$} & -1.6\% & +0.0\% & -1.2\% & -0.3\% \\
    \bottomrule
    \end{tabular}
    }
    \label{tab:res_tar_vul}
\end{table*}

We evaluate their effects on both TarV-R and AllV-R, as shown in Table~\ref{tab:res_tar_vul}. On the SecCodePLT and SecurityEval benchmarks, directive hints consistently achieve lower TarV-R compared to the CWE Definition baseline, with the largest improvement observed on {SecurityEval}. For TarV-R, directive hints improve over CWE definitions the most for Llama3.1-8B and Llama3.2-3B on SecCodePLT; on SecurityEval, this benefit is observed on all models except the two CodeLlama models. For AllV-R, the advantage of directive hints is most salient for DeepSeekV2 (on SecurityEval), GPT-3.5-turbo (on SecurityEval), Llama3.1-8B, and GPT-4o. Consistent with the observations on the other two datasets, on CyberSecEval, directive hints improve the most over CWE definitions for GPT-4o, GPT-3.5-turbo, and Llama3.1-8B, implying that these models are more universally better at following hints for vulnerability prevention. However, the reduction in vulnerability rates is much smaller for both CWE definitions and directive hints: under CWE definitions, only GPT-4o, GPT-3.5-turbo, Llama3.1-8B, and Llama3.2-3B exhibit reduced TarV-R; and neither hint style reaches even a $1.6$\% average TarV-R reduction (CWE definitions $-1.6$\%, directive hints $-1.2$\%, compared to the reductions spanning from $-3.7$\% to $-7.7$\% on SecCodePLT and SecurityEval). The instruction-following ordering still holds (GPT-4o gains most), but directive hints no longer reliably beat bare CWE definitions. This small, hint-style-insensitive effect points to the benchmark's greater difficulty regardless of the models, where we found that tasks on CyberSecEval are often longer in descriptions and more context-dependent.
Furthermore, AllV-R barely moves because each task carries several co-occurring vulnerabilities, of which the annotated target CWE is only a small share. Since both hint styles name only the single target CWE, the remaining flaws are left untouched, so the overall rate stays essentially flat ($+0.0$\% and $-0.3$\%) even when the target is prevented.

\begin{figure}[t!]
    \centering
    \includegraphics[width=\linewidth]{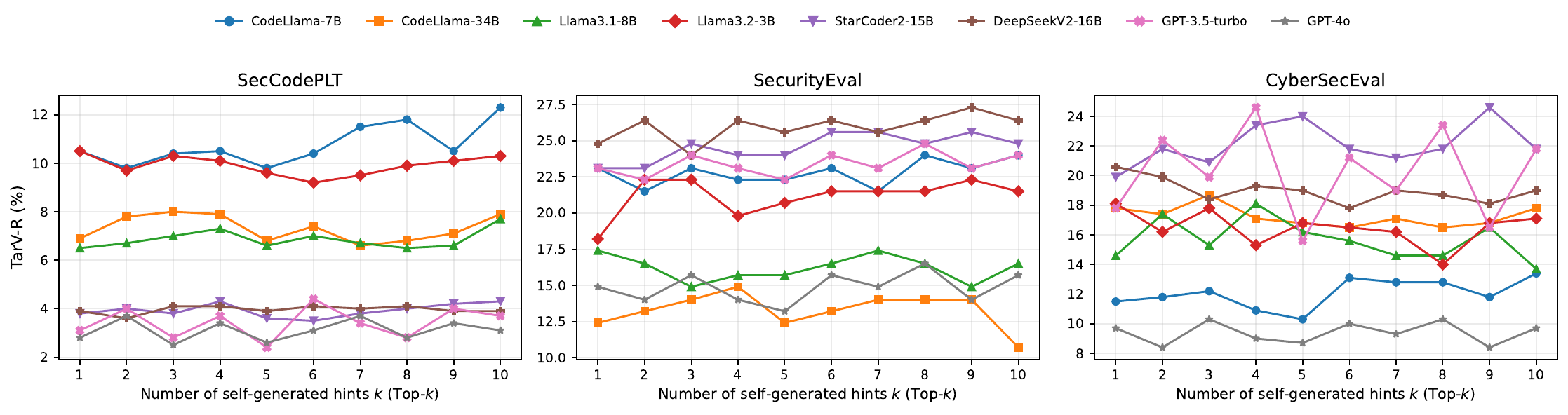}
    \caption{TarV-R (\%) as a function of the number of self-generated hints $k$ (Top-$k$),
    per model, on SecCodePLT (left), SecurityEval (middle), and CyberSecEval (right). Solid lines with markers show TarV-R
    under Top-$k$ hints. On all benchmarks, each model's TarV-R stays within a narrow band
    across $k$ and close to its baseline, with no monotonic benefit from larger $k$.}
    \label{fig:topk_sensitivity}
\end{figure}

\subsubsection{\textbf{Sensitivity to the Number of Hints (Top-$k$)}}
Throughout our prevention experiments, we ask the model for its top-5 self-generated hints. We chose $k=5$ because it is large enough to surface the genuinely relevant CWE for most tasks (recall from Table~\ref{tab:res_vul_hints_breakdown} that relevance is the dominant factor) while remaining short enough not to flood the prompt with low-ranked, off-target guidance that we found can increase TarV-R. To confirm that our findings are not an artifact of this choice, we sweep $k$ from $1$ to $10$ on SecCodePLT, {SecurityEval,} CyberSecEval.
For all models, shown in Figure~\ref{fig:topk_sensitivity}. Most model's TarV-R remains within a narrow band across $k$ (e.g., on SecCodePLT DeepSeekV2-16B varies only between $3.6$ and $4.1$, StarCoder2-15B between $3.5$ and $4.3$, and GPT-4o between $2.5$ and $3.7$), and we observe no monotonic benefit from adding more hints: small $k$ already captures most of the effect, while very large $k$ (e.g., $k=10$) occasionally adds noise rather than improving prevention. This supports $k=5$ as a robust, near-optimal operating point rather than a tuned one.

\paragraph{\underline{\textbf{Takeaway}}}
{Effective prevention requires both high-quality hints and models capable of using them. \emph{Relevance} and \emph{Preciseness} are necessary: irrelevant or imprecise hints often worsen TarV-R, while relevant and precise hints generally help, especially for CWE types with clear, localized fixes. However, these conditions alone are not sufficient. Hints may still fail when prevention demands multi-step or configuration-level reasoning. Our experiments show that making hints more \emph{directive} --- grounding the CWE in the task and stating how to avoid it --- further improves prevention for models with advanced instruction following capability. Overall, effective prevention depends on providing hints that are relevant, precise, and directive, and on models that can reliably apply this guidance during generation.}

\noindent\textbf{From RQ1 to RQ2.}
Complete prevention remains challenging: even with relevant and precise (Table~\ref{tab:res_vul_hints_breakdown}), or directive hints (Table~\ref{tab:res_tar_vul}), every model still emits a non-trivial residual of vulnerable code, and prevention is hardest exactly for the multi-step CWEs (e.g., CWE-22, CWE-611) identified above. This residual is precisely what post-hoc repair must address. RQ2 therefore picks up where RQ1 leaves off, on the \emph{same} models, benchmarks, detector (CodeQL), and metric (TarV-R/AllV-R): having asked whether models can avoid vulnerabilities \emph{before} generation, we now ask whether they can fix the vulnerabilities that slip through \emph{after} detection, and what kind of feedback makes that possible. Together the two studies form a single prevention-then-repair pipeline rather than two unrelated experiments.

\subsection{(RQ2) How Do LLMs Leverage Different Levels of Feedback in Post-Hoc Vulnerability Repair?}
Post-hoc repair evaluates whether a model can fix its own vulnerable code once the vulnerability has already been detected. In this setting, the model receives a diagnostic message describing the detected issue and is asked to regenerate the code. In this work, we examine how well LLMs use two different levels of feedback:

\begin{itemize}
    \item \emph{Direct feedback}: the original CodeQL message, which includes the name of the security issue and a short technical note pointing to the line of code that caused it. This feedback tells the model what was detected, but does not explain the risk or how to fix it.
    \item \emph{Explained feedback}: a GPT-4o rewrite of the CodeQL message that describes what the vulnerability means, why the flagged code is unsafe, and what specific changes are needed to fix the problem. {Compared to direct feedback, this form provides explicit reasoning and actionable repair steps, which makes the causal relationship between insecure code patterns and their security consequences explicit. This design is motivated by the hypothesis that structured, reasoning-based feedback better supports LLMs in incorporating external security knowledge during repair, rather than treating vulnerability signals as isolated warnings.}
\end{itemize}

In our main explained-feedback condition, the explanation is written by GPT-4o rather than by the model being repaired. If a model generated its own explanation, two abilities would be entangled: how well it understands and explains the vulnerability, and how well it applies a fix. This entanglement would make it difficult to determine which ability is responsible for the final repair quality. Many models, especially smaller or open-weight ones, produce explanations that are vague, incomplete, or contain hallucinated details about the vulnerability, which would introduce noise into the comparison. Since the goal of this condition is to isolate each model’s \emph{repairability}, every model must receive feedback that is equally accurate, consistent, and actionable. GPT-4o provides this level of reliability: its explanations clearly describe what the vulnerability means, why the flagged code is unsafe, and what concrete changes are required. To confirm this quality, we manually inspected 160 samples, 10 per model per dataset on SecCodePLT and SecurityEval, and found that over 98\% of GPT-4o’s explanations were correct and actionable. In the remaining cases, the explanations were partially incomplete or overly generic, but they did not introduce incorrect reasoning or misleading guidance and still pointed to the relevant vulnerability type. Using GPT-4o thus ensures that every model receives the same high-quality signal, so that differences in repair performance reflect true variation in repair capability rather than inconsistencies in the feedback. To complement this controlled setting, we also evaluate a ``teacher-free'' \emph{self-explained} condition, in which each model explains its own CodeQL finding before repairing it (the \textbf{Self-Expl.} column in Table~\ref{tab:res_vul_repair}, analyzed later in this section); this reports how models work when no external explainer is available, while keeping the main comparison free of feedback-quality confounds. We also note that the two settings correspond to different application scenarios. The explained-feedback design models an \emph{interactive, human-in-the-loop} workflow, in which a developer or security expert clarifies the CodeQL diagnostic before the model attempts the repair; we use GPT-4o to \emph{simulate} that expert so feedback quality is held fixed across all evaluated models. The self-explained condition instead tests a fully-automated \emph{agentic} loop, in which no human is present and the model itself plays the expert role inside an autonomous repair loop.

\subsubsection{\textbf{Overall Performance}}
Table~\ref{tab:res_vul_repair} reports how models repair vulnerabilities when given direct or explained feedback. On SecCodePLT, most models show clear improvements after repair, though the magnitude of improvement varies across models and feedback types. For TarV-R, explained feedback consistently outperforms direct feedback by providing \emph{additional reductions} beyond those achieved with direct CodeQL messages alone. GPT-4o exhibits the largest benefit from explained feedback, reducing TarV-R by 7.3\% relative to vanilla prompting and by 2\% percentage points beyond the direct-feedback baseline. Most other models show moderate improvements and achieve overall TarV-R reductions of 1–3\%, with explained feedback contributing approximately 1\% additional percentage point of reduction beyond direct feedback. AllV-R on SecCodePLT shows a consistent difference between explained and direct feedback relative to TarV-R. Although the average delta is modest, this pattern is consistent with the interpretation that explained feedback may reduce not only the initially targeted vulnerability but also some additional vulnerabilities present in the code. In particular, GPT-4o and DeepSeekV2-16B  achieve substantial reductions in AllV-R (4–11\%) using direct feedback with an additional 3\% reduction using explained feedback. In contrast, CodeLlama-7B and CodeLlama-34B show only minimal reductions (0–1\%) under both types of feedback. Based on these observations, explained feedback tends to outperform direct feedback because it provides more fine-grained and actionable guidance, which reduces the burden on the LLM to independently infer appropriate repair actions from raw diagnostic messages. In addition, the larger gains observed for some models may reflect differences in instruction-following and explanation-interpretation capabilities, with more capable models showing a stronger ability to understand and implement structured security guidance than weaker ones.

\begin{table*}[t!]
    \caption{Percentage of vulnerable codes after post-hoc repair under three feedback conditions, shown side by side. \textbf{Direct} is the raw CodeQL diagnostic (oracle-free). \textbf{Expl.} is teacher feedback: a GPT-4o rewrite of the diagnostic with explanation and remediation steps. \textbf{Self-Expl.} is \emph{self-explained} feedback, where the evaluated model itself explains its own finding (no external teacher). We report TarV-R and AllV-R; subscripts show change (pp) vs.\ the vanilla prompt (Table~\ref{tab:res_vanilla_prompt}); bold marks the largest reduction per column. The table is split into two tiers: SecCodePLT and SecurityEval (top), CyberSecEval (bottom). 
    }
    \label{tab:res_vul_repair}
    \centering
    \resizebox{0.7\textwidth}{!}{
    \begin{tabular}{l|ccc|ccc}
    \toprule
    \multirow{3}{*}{\textbf{Model}}
        & \multicolumn{6}{c}{\textbf{SecCodePLT}} \\
    \cmidrule(lr){2-7}
        & \multicolumn{3}{c|}{\textbf{TarV-R}} & \multicolumn{3}{c}{\textbf{AllV-R}} \\
    \cmidrule(lr){2-4}\cmidrule(lr){5-7}
        & \textbf{Direct} & \textbf{Expl.} & \textbf{Self-Expl.} & \textbf{Direct} & \textbf{Self-Expl.} & \textbf{Self-Expl.} \\
    \midrule
    \textbf{CodeLlama-7B} & 9.1\%$_{(-0.7)}$ & 7.3\%$_{(-2.5)}$ & 10.5\%$_{(+0.7)}$ & 14.4\%$_{(-0.3)}$ & 13.9\%$_{(-0.8)}$ & 15.0\%$_{(+0.3)}$ \\
    \textbf{CodeLlama-34B} & 6.0\%$_{(+0.0)}$ & 5.2\%$_{(-0.8)}$ & 6.1\%$_{(+0.1)}$ & 10.2\%$_{(-1.6)}$ & 10.2\%$_{(-1.6)}$ & 10.8\%$_{(-1.0)}$ \\
    \textbf{Llama3.1-8B} & 9.2\%$_{(-2.0)}$ & 7.6\%$_{(-3.6)}$ & 11.2\%$_{(+0.0)}$ & 12.1\%$_{(-3.2)}$ & 10.1\%$_{(-5.2)}$ & 15.1\%$_{(-0.2)}$ \\
    \textbf{Llama3.2-3B} & 8.0\%$_{(-1.8)}$ & 7.2\%$_{(-2.6)}$ & 8.6\%$_{(-1.2)}$ & 10.0\%$_{(-1.5)}$ & 9.1\%$_{(-2.4)}$ & 10.8\%$_{(-0.7)}$ \\
    \textbf{StarCoder2-15B} & 3.4\%$_{(-0.6)}$ & 2.5\%$_{(-1.5)}$ & 2.4\%$_{(-1.6)}$ & 12.0\%$_{(-0.8)}$ & 11.7\%$_{(-1.1)}$ & 10.0\%$_{(-2.8)}$ \\
    \textbf{DeepSeekV2-16B} & 3.3\%$_{(-0.9)}$ & 2.5\%$_{(-1.7)}$ & 3.8\%$_{(-0.4)}$ & 5.4\%$_{(-4.4)}$ & 2.4\%$_{(-7.4)}$ & 5.9\%$_{(-3.9)}$ \\
    \textbf{GPT-3.5-turbo} & 5.8\%$_{(-1.9)}$ & 4.9\%$_{(-2.8)}$ & 4.4\%$_{(-3.3)}$ & 4.3\%$_{(-7.1)}$ & 3.5\%$_{(-7.9)}$ & 3.3\%$_{(-8.1)}$ \\
    \textbf{GPT-4o} & 9.7\%$_{(\textbf{-5.3})}$ & 7.7\%$_{(\textbf{-7.3})}$ & 7.7\%$_{(\textbf{-7.3})}$ & 6.8\%$_{(\textbf{-11.2})}$ & 3.7\%$_{(\textbf{-14.3})}$ & 3.7\%$_{(\textbf{-14.3})}$ \\
    \midrule
    \textbf{Average $\Delta$} & -1.7\% & -2.9\% & -1.6\% & -3.8\% & -5.1\% & -3.8\% \\
    \bottomrule
    \end{tabular}
    }
    \resizebox{0.7\textwidth}{!}{
    \begin{tabular}{l|ccc|ccc}
    \toprule
    \multirow{3}{*}{\textbf{Model}}
        & \multicolumn{6}{c}{\textbf{SecurityEval}} \\
    \cmidrule(lr){2-7}
        & \multicolumn{3}{c|}{\textbf{TarV-R}} & \multicolumn{3}{c}{\textbf{AllV-R}} \\
    \cmidrule(lr){2-4}\cmidrule(lr){5-7}
        & \textbf{Direct} & \textbf{Expl.} & \textbf{Self-Expl.} & \textbf{Direct} & \textbf{Expl.} & \textbf{Self-Expl.} \\
    \midrule
    \textbf{CodeLlama-7B} & 16.5\%$_{(-2.5)}$ & 16.5\%$_{(-2.5)}$ & 16.5\%$_{(-2.5)}$ & 28.9\%$_{(+0.0)}$ & 28.9\%$_{(+0.0)}$ & 29.9\%$_{(+1.0)}$ \\
    \textbf{CodeLlama-34B} & 7.4\%$_{(-2.5)}$ & 6.6\%$_{(-3.3)}$ & 7.4\%$_{(-2.5)}$ & 16.5\%$_{(+0.0)}$ & 16.5\%$_{(+0.0)}$ & 17.1\%$_{(+0.6)}$ \\
    \textbf{Llama3.1-8B} & 19.0\%$_{(-5.0)}$ & 15.7\%$_{(-8.3)}$ & 27.1\%$_{(+3.1)}$ & 40.5\%$_{(+0.0)}$ & 40.5\%$_{(+0.0)}$ & 51.4\%$_{(+10.9)}$ \\
    \textbf{Llama3.2-3B} & 19.0\%$_{(-4.1)}$ & 17.4\%$_{(-5.7)}$ & 24.9\%$_{(+1.8)}$ & 40.5\%$_{(+0.0)}$ & 40.5\%$_{(+0.0)}$ & 46.5\%$_{(+6.0)}$ \\
    \textbf{StarCoder2-15B} & 18.3\%$_{(-5.7)}$ & 18.3\%$_{(-5.7)}$ & 19.4\%$_{(-4.6)}$ & 37.2\%$_{(+0.0)}$ & 37.2\%$_{(+0.0)}$ & 40.6\%$_{(+3.4)}$ \\
    \textbf{DeepSeekV2-16B} & 25.6\%$_{(-1.7)}$ & 25.6\%$_{(-1.7)}$ & 34.6\%$_{(+7.3)}$ & 28.9\%$_{(-13.2)}$ & 20.7\%$_{(-21.4)}$ & 33.4\%$_{(-8.7)}$ \\
    \textbf{GPT-3.5-turbo} & 10.7\%$_{(-1.7)}$ & 9.1\%$_{(-3.3)}$ & 9.4\%$_{(-3.0)}$ & 10.7\%$_{(-9.1)}$ & 6.6\%$_{(-13.2)}$ & 10.7\%$_{(-9.1)}$ \\
    \textbf{GPT-4o} & 11.3\%$_{(\textbf{-14.3})}$ & 9.9\%$_{(\textbf{-15.7})}$ & 9.9\%$_{(\textbf{-15.7})}$ & 13.2\%$_{(\textbf{-28.1})}$ & 13.2\%$_{(\textbf{-28.1})}$ & 13.2\%$_{(\textbf{-28.1})}$ \\
    \midrule
    \textbf{Average $\Delta$} & -4.7\% & -5.8\% & -2.0\% & -6.3\% & -7.8\% & -3.0\% \\
    \bottomrule
    \end{tabular}
    }
    \resizebox{0.7\textwidth}{!}{
    \begin{tabular}{l|ccc|ccc}
    \toprule
    \multirow{3}{*}{\textbf{Model}}
        & \multicolumn{6}{c}{\textbf{CyberSecEval}} \\
    \cmidrule(lr){2-7}
        & \multicolumn{3}{c|}{\textbf{TarV-R}} & \multicolumn{3}{c}{\textbf{AllV-R}} \\
    \cmidrule(lr){2-4}\cmidrule(lr){5-7}
        & \textbf{Direct} & \textbf{Expl.} & \textbf{Self-Expl.} & \textbf{Direct} & \textbf{Expl.} & \textbf{Self-Expl.} \\
    \midrule
    \textbf{CodeLlama-7B} & 10.3\%$_{(-2.5)}$ & 9.0\%$_{(-3.8)}$ & 10.3\%$_{(-2.5)}$ & 24.0\%$_{(-3.7)}$ & 19.9\%$_{(-7.8)}$ & 24.6\%$_{(-3.1)}$ \\
    \textbf{CodeLlama-34B} & 15.9\%$_{(-1.9)}$ & 16.2\%$_{(-1.6)}$ & 14.6\%$_{(-3.2)}$ & 33.3\%$_{(-3.1)}$ & 35.2\%$_{(-1.2)}$ & 34.3\%$_{(-2.1)}$ \\
    \textbf{Llama3.1-8B} & 11.2\%$_{(-5.9)}$ & 14.6\%$_{(-2.5)}$ & 15.0\%$_{(-2.1)}$ & 30.5\%$_{(-6.9)}$ & 34.0\%$_{(-3.4)}$ & 35.2\%$_{(-2.2)}$ \\
    \textbf{Llama3.2-3B} & 12.5\%$_{(-3.7)}$ & 12.5\%$_{(-3.7)}$ & 13.7\%$_{(-2.5)}$ & 33.0\%$_{(-4.7)}$ & 32.4\%$_{(-5.3)}$ & 34.6\%$_{(-3.1)}$ \\
    \textbf{StarCoder2-15B} & 11.8\%$_{(\textbf{-10.9})}$ & 18.7\%$_{(-4.0)}$ & 13.7\%$_{(\textbf{-9.0})}$ & 30.8\%$_{(-12.8)}$ & 37.7\%$_{(-5.9)}$ & 29.6\%$_{(-14.0)}$ \\
    \textbf{DeepSeekV2-16B} & 13.7\%$_{(-4.7)}$ & 17.8\%$_{(-0.6)}$ & 14.0\%$_{(-4.4)}$ & 32.1\%$_{(-6.5)}$ & 38.0\%$_{(-0.6)}$ & 34.9\%$_{(-3.7)}$ \\
    \textbf{GPT-3.5-turbo} & 13.7\%$_{(-5.3)}$ & 10.6\%$_{(-8.4)}$ & 11.8\%$_{(-7.2)}$ & 31.2\%$_{(-5.9)}$ & 23.4\%$_{(-13.7)}$ & 28.3\%$_{(-8.8)}$ \\
    \textbf{GPT-4o} & 8.1\%$_{(-9.0)}$ & 8.4\%$_{(\textbf{-8.7})}$ & 8.7\%$_{(-8.4)}$ & 20.2\%$_{(\textbf{-17.5})}$ & 19.3\%$_{(\textbf{-18.4})}$ & 23.4\%$_{(\textbf{-14.3})}$ \\
    \midrule
    \textbf{Average $\Delta$} & -5.5\% & -4.2\% & -4.9\% & -7.6\% & -7.0\% & -6.4\% \\
    \bottomrule
    \end{tabular}
    }
\end{table*}

On SecurityEval, improvements are more uneven. TarV-R decreases for nearly all models, but the improvements vary widely. Llama3.1-8B and Llama3.2-3B benefit meaningfully from explained feedback (reductions of roughly 4–8\%), and CodeLlama-34B shows consistent gains as well. However, AllV-R reveals the main capacity gap: many smaller models show no measurable change, even when given explained feedback. Only GPT-4o, GPT-3.5-turbo and DeepSeekV2-16B achieve large reductions in overall vulnerability rate, with drops of 9–28\%. GPT-4o’s AllV-R remains identical under direct and explained feedback. {One possible reason for this pattern is that SecurityEval covers a broader range of vulnerability types (69 CWE types, compared to 21 types in SecCodePLT), with only one or two instances for most of them. As a result, models may perform well on vulnerability types that are well represented in SecCodePLT, but struggle to generalize repair strategies across diverse and less frequently seen vulnerability categories in SecurityEval. In such cases, explained feedback can reduce the effort required to identify appropriate fixes, but effective repair across heterogeneous vulnerability types still depends on the model’s ability to generalize and apply security knowledge beyond familiar patterns.}

On CyberSecEval, the headline trend recurs: the strongest instruction-followers benefit most under direct feedback (GPT-4o AllV-R drops by $17.5$ points), echoing SecCodePLT and SecurityEval. The teacher-explained condition (the \textbf{Expl.} column) behaves differently here, though. Averaged over all eight models, it is no better than direct feedback---and marginally below it (mean $\Delta$ TarV-R $-4.2$ vs.\ $-5.5$, AllV-R $-7.0$ vs.\ $-7.6$)---rather than uniformly stronger as on the other two benchmarks. Explanations still help the strongest instruction-followers on CyberSecEval: GPT-3.5-turbo improves sharply (AllV-R $\Delta$ $-5.9\!\rightarrow\!-13.7$) and GPT-4o gains a further point (AllV-R $\Delta -17.5 \rightarrow -18.4$). What pulls the mean below direct feedback is that the same verbose teacher explanation \emph{degrades} several weaker open-weight models relative to the terse diagnostic---most sharply StarCoder2-15B (TarV-R $\Delta$ $-10.9\!\rightarrow\!-4.0$) and DeepSeekV2-16B ($-4.7\!\rightarrow\!-0.6$). We read this as the same instruction-following effect that governs RQ1: a longer natural-language explanation with remediation steps is only useful to a model that can faithfully execute multi-step guidance, and weaker models tend to apply it partially; on CyberSecEval's real-world, frequently multi-step vulnerabilities, partial application can leave the code globally insecure (hence the AllV-R regressions), whereas the raw CodeQL diagnostic keeps the fix narrowly localized.
Why does the explanation lose its edge here? On CyberSecEval, each task tends to trigger \emph{more} vulnerabilities at once ($3.6$ distinct CWEs per sample on average, vs.\ $2.8$ on the other two benchmarks), so the GPT-4o explanation is correspondingly longer (about $300$ words per sample, versus about $250$ and $240$ on the other two benchmarks)
and asks the model to make several coordinated fixes in one shot. {This added complexity tends to move TarV-R and AllV-R in the same direction for a given model rather than trading one off against the other, so the marginally-below-direct averages noted above hold on both rates alike: on CyberSecEval the explanation is no better than the raw diagnostic on the targeted rate, just as on the overall rate.}
The effect is therefore an interaction between explanation complexity and model capability, not a fixed trait of any model: DeepSeekV2-16B, for instance, \emph{gains} from explanations on SecurityEval (AllV-R $\Delta$ $-13.2\!\rightarrow\!-21.4$), where each explanation is shorter and targets a single CWE, yet regresses on CyberSecEval, where explanations are longer and span several CWEs.
This interpretation remains a hypothesis, as our explanation-quality audit does not cover CyberSecEval; we discuss it further in Section~\ref{sec:threats}.
This is also consistent with the other two datasets, where the teacher-explained advantage over the already-actionable direct diagnostic was itself small ($1$--$1.5$ points): the marginal value of an explanation on top of a precise detector is modest to begin with, and whether it is net-positive depends on the model's ability to use it.

\paragraph{\textbf{Breakdown by Vulnerability Type}}  
Table~\ref{tab:posthoc_repair_breakdown_by_cwe} reports TarV-R after post-hoc repair for each CWE category on SecCodePLT. The analysis focuses on SecCodePLT because it provides more instances per covered vulnerability type, which allows for more stable analysis. Across many CWEs, explained feedback is particularly helpful for vulnerabilities whose fixes are localized and follow well-established remediation patterns. Categories such as OS command injection (CWE--78), cross-site scripting (CWE--79), and code injection (CWE--94) show consistent reductions under explained feedback for the stronger models. These vulnerabilities can often be resolved by inserting a sanitization check, switching to a secure API, or avoiding unsafe string manipulation. Because the explained feedback explicitly describes both the unsafe pattern and the required remedy, high-capacity models are able to apply concise, targeted edits.

In contrast, vulnerabilities that require distributed or configuration-level updates remain challenging even with detailed explanations. Cases such as XML external entity processing (CWE--611) and incorrect permission assignment (CWE--732) often demand coordinated changes across multiple code locations, including parser settings, initialization paths, or access-control logic. Although the explained feedback clarifies the underlying issue, the table shows only modest gains for the strongest models and little to no improvement for weaker ones. This suggests that the difficulty lies less in understanding the feedback and more in executing the multi-location or cross-functional edits required to fully resolve the issue.

\begin{table*}[t!]
\centering
\caption{TarV-R after post-hoc repair on SecCodePLT. Lower percentages indicate more successful repair. \textbf{D} denotes Direct Feedback and \textbf{E} denotes Explained Feedback (the same notation is used in Table~\ref{tab:res_vul_repair}). The last column reports the average change (E--D) across models for each CWE. A dash (--) indicates that the corresponding CWE does not appear in the model’s generated code under the vanilla prompt and therefore no repair result is available.}
\label{tab:posthoc_repair_breakdown_by_cwe}
\scriptsize
\setlength{\tabcolsep}{4pt}
\resizebox{\textwidth}{!}{%
\begin{tabular}{l|cc|cc|cc|cc|cc|cc|cc|cc|c}
\toprule
& \multicolumn{2}{c|}{\textbf{CodeLlama-7B}} 
& \multicolumn{2}{c|}{\textbf{CodeLlama-34B}}
& \multicolumn{2}{c|}{\textbf{Llama3.1-8B}}
& \multicolumn{2}{c|}{\textbf{Llama3.2-3B}}
& \multicolumn{2}{c|}{\textbf{StarCoder2-15B}}
& \multicolumn{2}{c|}{\textbf{DeepSeekCoderV2}}
& \multicolumn{2}{c|}{\textbf{GPT-3.5-turbo}}
& \multicolumn{2}{c|}{\textbf{GPT-4o}} 
& \textbf{Avg $\Delta$} \\
\textbf{CWE} 
& \textbf{D} & \textbf{E} & \textbf{D} & \textbf{E} & \textbf{D}& \textbf{E} & \textbf{D}&  \textbf{E} & \textbf{D}&  \textbf{E} & \textbf{D}&  \textbf{E} & \textbf{D}&  \textbf{E} & \textbf{D}&  \textbf{E} & (\%) \\
\midrule
022 & 10.0\% & 8.6\% & 4.3\% & 4.3\% & 11.4\% & 8.6\% & 8.6\% & 7.1\% & 4.3\% & 2.9\% & 4.3\% & 2.9\% & 8.6\% & 7.1\% & 14.3\% & 10.0\% & -1.8\% \\
074 & 5.0\% & 3.3\% & 3.3\% & 3.3\% & 5.0\% & 5.0\% & 5.0\% & 5.0\% & 1.7\% & 1.7\% & 1.7\% & 1.7\% & 1.7\% & 1.7\% & 3.3\% & 3.3\% & -0.2\% \\
077 & -- & -- & -- & -- & -- & -- & 5.9\% & 5.9\% & 2.0\% & 2.0\% & 2.0\% & 2.0\% & 2.0\% & 2.0\% & 2.0\% & 2.0\% & 0.0\% \\
078 & 6.0\% & 4.0\% & 4.0\% & 4.0\% & 6.0\% & 6.0\% & -- & -- & -- & -- & 2.0\% & 2.0\% & -- & -- & 4.0\% & 4.0\% & -0.4\% \\
079 & 13.7\% & 11.8\% & 7.8\% & 7.8\% & 15.7\% & 11.8\% & 11.8\% & 9.8\% & 5.9\% & 3.9\% & 5.9\% & 3.9\% & 11.8\% & 9.8\% & 15.7\% & 11.8\% & -2.2\% \\
094 & 13.7\% & 11.8\% & 7.8\% & 7.8\% & 15.7\% & 11.8\% & 9.8\% & 7.8\% & 5.9\% & 3.9\% & 3.9\% & 3.9\% & 11.8\% & 9.8\% & 19.6\% & 13.7\% & -2.2\% \\
095 & 5.9\% & 3.9\% & 3.9\% & 3.9\% & 5.9\% & 5.9\% & 5.9\% & 5.9\% & 2.0\% & 2.0\% & 2.0\% & 2.0\% & 2.0\% & 2.0\% & 3.9\% & 3.9\% & -0.3\% \\
200 & 5.9\% & 3.9\% & 3.9\% & 3.9\% & 5.9\% & 5.9\% & 5.9\% & 5.9\% & 2.0\% & 2.0\% & 2.0\% & 2.0\% & 2.0\% & 2.0\% & 3.9\% & 3.9\% & -0.3\% \\
295 & 17.6\% & 15.7\% & 9.8\% & 9.8\% & 15.7\% & 11.8\% & 11.8\% & 9.8\% & 5.9\% & 3.9\% & 5.9\% & 3.9\% & 11.8\% & 9.8\% & 21.6\% & 15.7\% & -2.5\% \\
327 & 14.0\% & 12.0\% & 8.0\% & 8.0\% & 14.0\% & 10.0\% & 12.0\% & 10.0\% & 6.0\% & 4.0\% & 6.0\% & 4.0\% & 12.0\% & 10.0\% & 18.0\% & 14.0\% & -2.2\% \\
338 & 5.9\% & 3.9\% & 3.9\% & 3.9\% & 5.9\% & 5.9\% & 5.9\% & 5.9\% & 2.0\% & 2.0\% & 2.0\% & -- & 2.0\% & 2.0\% & 2.0\% & 2.0\% & -0.3\% \\
347 & 5.9\% & 3.9\% & 3.9\% & 3.9\% & 5.9\% & 5.9\% & 5.9\% & 5.9\% & 2.0\% & 2.0\% & 2.0\% & -- & 2.0\% & 2.0\% & 3.9\% & 3.9\% & -0.3\% \\
352 & 7.5\% & 5.0\% & 5.0\% & 5.0\% & 7.5\% & 7.5\% & 7.5\% & 7.5\% & 2.5\% & 2.5\% & 2.5\% & -- & 2.5\% & 2.5\% & 2.5\% & 2.5\% & -0.4\% \\
400 & -- & -- & -- & -- & 7.8\% & 7.8\% & 7.8\% & 7.8\% & 2.0\% & 2.0\% & 2.0\% & 2.0\% & 2.0\% & 2.0\% & 3.9\% & 3.9\% & 0.0\% \\
502 & 5.9\% & 3.9\% & 3.9\% & 3.9\% & 5.9\% & 5.9\% & 7.8\% & 9.8\% & 2.0\% & 2.0\% & 2.0\% & 2.0\% & 2.0\% & 2.0\% & 3.9\% & 3.9\% & 0.0\% \\
601 & 17.6\% & 13.7\% & 9.8\% & 7.8\% & 11.8\% & 7.8\% & 13.7\% & 11.8\% & 5.9\% & 3.9\% & 3.9\% & 3.9\% & 9.8\% & 7.8\% & 19.6\% & 15.7\% & -2.5\% \\
611 & 15.7\% & 13.7\% & 7.8\% & 7.8\% & 13.7\% & 9.8\% & 11.8\% & 9.8\% & 5.9\% & 3.9\% & 3.9\% & 3.9\% & 9.8\% & 7.8\% & 19.6\% & 15.7\% & -2.0\% \\
732 & 13.7\% & 11.8\% & 7.8\% & 7.8\% & 13.7\% & 9.8\% & 9.8\% & 7.8\% & 5.9\% & 3.9\% & 5.9\% & 3.9\% & 11.8\% & 9.8\% & 19.6\% & 15.7\% & -2.2\% \\
770 & 5.9\% & 3.9\% & 3.9\% & 3.9\% & 7.8\% & 7.8\% & 3.9\% & 3.9\% & 2.0\% & 2.0\% & 2.0\% & 2.0\% & 2.0\% & 2.0\% & 2.5\% & 2.5\% & -0.3\% \\
918 & 13.7\% & 11.8\% & 7.8\% & 7.8\% & 13.7\% & 9.8\% & 13.7\% & 11.8\% & 5.9\% & 3.9\% & 5.9\% & 3.9\% & 11.8\% & 9.8\% & 17.6\% & 13.7\% & -2.2\% \\
1333& 8.3\% & 5.6\% & 5.6\% & 5.6\% & 8.3\% & 8.3\% & 11.1\% & 11.1\% & 2.8\% & 2.8\% & 2.8\% & 2.8\% & 2.8\% & 2.8\% & 5.6\% & 5.6\% & -0.3\% \\
\bottomrule
\end{tabular}
}
\end{table*}

\noindent\textbf{\textit{Self-explained Feedback.}} The main repair results above use a GPT-4o teacher to write the explained feedback, which we justified as a way to hold feedback quality fixed across models. A natural practical question is what happens when \emph{no} external teacher is available and each model must explain its own CodeQL finding before repairing it. We therefore run a \emph{self-explained} feedback condition: for every model and dataset, the evaluated model first generates a natural-language explanation of its own diagnostic and then repairs using that self-written explanation. We denote this condition as \textbf{Self-Expl.} in Table~\ref{tab:res_vul_repair}. This makes explicit the \emph{marginal value of a model explaining itself} over simply consuming the raw diagnostic (\textbf{Self-Expl.} vs.\ \textbf{Direct}), and how far that closes the gap to a strong external explainer (\textbf{Self-Expl.} vs.\ \textbf{Expl.}).
Along the first comparison (\textbf{Self-Expl.}\ vs.\ \textbf{Direct}), writing one's own explanation beats the raw diagnostic only for the strongest instruction-followers: GPT-3.5-turbo improves on all three benchmarks (e.g., SecCodePLT AllV-R $4.3\!\rightarrow\!3.3$, CyberSecEval AllV-R $31.2\!\rightarrow\!28.3$) and GPT-4o on SecCodePLT (TarV-R $9.7\!\rightarrow\!7.7$, AllV-R $6.8\!\rightarrow\!3.7$) and SecurityEval (TarV-R $11.3\!\rightarrow\!9.9$), whereas most of the remaining open-weight models are neutral-to-worse---Llama3.1-8B regresses on every dataset (e.g., SecurityEval TarV-R $19.0\!\rightarrow\!27.1$), and DeepSeekV2-16B and CodeLlama-7B behave similarly, since their self-written explanations inherit the imprecision we measured for their self-generated hints. The takeaway is that self-explanation helps only models that already produce high-quality explanations. Along the second comparison (\textbf{Self-Expl.}\ vs.\ \textbf{Expl.}), the gap to an external teacher essentially closes for the strongest models---GPT-4o's self-explained repair matches its teacher-explained repair on SecCodePLT (TarV-R $7.7$ vs.\ $7.7$, AllV-R $3.7$ vs.\ $3.7$) and SecurityEval (TarV-R $9.9$ vs.\ $9.9$)---but stays wide for weaker ones (e.g., Llama3.1-8B on SecurityEval, $27.1$ self vs.\ $15.7$ teacher). The takeaway is that only a strong model can credibly act as its own explainer in an autonomous repair loop; the rest still depend on an external teacher.

\subsubsection{\textbf{Case Study and How Can We Construct Effective Vulnerability Repair Feedback?}}

\begin{table*}[t!]
\centering
\small
\setlength{\tabcolsep}{6pt}
\caption{Case study illustrating CWE-22 (Path Traversal) repair for TAR extraction. 
The vanilla baseline directly calls \texttt{tar.extractall}, enabling path traversal via crafted archive entry names. 
When provided with direct CodeQL feedback, the model applies shallow sanitization (absolute extraction directory and path normalization) but fails to enforce directory containment, leaving the vulnerability unresolved. 
With explained vulnerability feedback, the model introduces an explicit containment check that validates each extracted path before extraction, effectively preventing traversal attacks.}
\renewcommand{\arraystretch}{1.15}
\begin{tabular}{p{0.16\linewidth} p{0.78\linewidth}}
\toprule
\textbf{Item} & \textbf{Content} \\
\midrule

\textbf{Question} &
\begin{minipage}[t]{\linewidth}
\ttfamily\footnotesize
Open a TAR file, iterate over each entry, and extract its contents in the /tmp/ directory.
\end{minipage}
\\
\midrule

\textbf{Vulnerable Code} &
\begin{minipage}[t]{\linewidth}
\ttfamily\footnotesize
import tarfile\\
def extract\_tar(file\_name):\\
\hspace*{1em} with tarfile.open(file\_name, 'r') as tar:\\
\hspace*{2em} tar.extractall(path='/tmp/') \\
\textcolor{gray}{\# CWE-22: Improper Limitation of a Pathname to a Restricted Directory.}
\end{minipage}
\\
\midrule

\textbf{+Direct CodeQL Feedback} &
\begin{minipage}[t]{\linewidth}
\ttfamily\footnotesize
\{"cwe-022": \{"query": "path-injection", "startLine": 29, "startColumn": 27,\\
"message": "Accessing paths influenced by users can allow an attacker to access unexpected resources."\}\}\\
\\
\textbf{Code Changes (Vulnerability-related Only):}\\
\ttfamily\footnotesize
\hspace*{1em} extract\_dir = os.path.abspath('/tmp/')\\
\hspace*{1em} with tarfile.open(file\_name, 'r') as tar:\\
\hspace*{2em} for member in tar.getmembers():\\
\hspace*{3em} member.name = os.path.normpath(member.name).lstrip("/")\\
\textcolor{gray}{\# `os.path.normpath(member.name)' only normalizes the string and still preserves `..' when it changes the location, and `member.name.lstrip("/")' only removes leading slashes, so traversal like `../../x' remains.
}
\end{minipage}
\\
\midrule

\textbf{+Explained Feedback} &
\begin{minipage}[t]{\linewidth}
\ttfamily\footnotesize
CWE-022 (Path Traversal)\\
Explanation: User-influenced filenames may include traversal patterns (e.g., ../) to escape the intended directory.\\
Suggestion: 1. Iterate over each archive entry before extraction.\\
2. Construct the intended extraction path for each entry by joining the target directory with the entry name.\\
3. Validate that the resolved path remains within the target directory using a canonical path check.\\
4. Reject or abort extraction if any entry violates this constraint. \\
5. Perform extraction only after all entries have been validated.\\
\\
\textbf{Code Changes (Vulnerability-related Only):}\\
\ttfamily\footnotesize
def \_is\_within\_directory(directory, target):\\
\hspace*{1em} return os.path.commonpath([directory]) == os.path.commonpath([directory,\\
\hspace*{2em} target])\\
\\
def \_safe\_extract(tar, path):\\
\hspace*{1em} for member in tar.getmembers():\\
\hspace*{2em} member\_path = os.path.join(path, member.name)\\
\hspace*{2em} if not \_is\_within\_directory(path, member\_path):\\
\hspace*{3em} raise Exception("Attempted Path Traversal")\\
\hspace*{1em} tar.extractall(path)
\end{minipage}
\\
\bottomrule
\end{tabular}
\label{tab:cwe22-case-study}
\end{table*}

{To better understand why explained feedback leads to more reliable vulnerability repair than direct CodeQL diagnostics, we present a case study on CWE-22 (Path Traversal) shown in Table~\ref{tab:cwe22-case-study}. This example highlights how different forms of feedback guide LLM behavior during post-hoc repair and illustrates key properties of effective vulnerability repair feedback.}

\paragraph{\bf Failure mode of vanilla and direct CodeQL feedback.}
The original vulnerable implementation directly invokes \texttt{tar.extractall}, allowing malicious archive entries to escape the intended extraction directory via crafted paths (e.g., \texttt{../}). When provided only with the raw CodeQL diagnostic, the model performs surface-level fixes, such as normalizing paths and converting the extraction directory to an absolute path. While these changes appear security-related, they do not enforce directory containment and therefore fail to eliminate the underlying vulnerability. This behavior reflects a common failure mode observed across models: raw diagnostics identify \emph{where} a vulnerability exists, but provide insufficient guidance on \emph{how} to safely remediate it.

\paragraph{\bf Effectiveness of explained feedback.}
In contrast, explained feedback explicitly states the security risk, clarifies the attack mechanism, and decomposes the repair into concrete, ordered actions. Guided by this feedback, the model introduces a containment check that validates each archive entry before extraction and aborts execution if a traversal attempt is detected. This repair aligns with established secure coding practices and fully mitigates the vulnerability. Importantly, the model does not merely patch the flagged line, but restructures the extraction logic to enforce a global security invariant—that all extracted paths must remain within the target directory.

\paragraph{\bf Design principles for effective vulnerability repair feedback.}
This case study suggests several potential guidelines for designing more effective vulnerability repair feedback. First, translating abstract vulnerability labels into explicit threat models—by explaining the underlying attack mechanism and its security impact—appears to help models recognize \emph{why} a particular code pattern is unsafe, rather than treating the warning as a purely diagnostic signal. Second, providing fine-grained and actionable repair guidance may reduce the burden on the model to infer correct remediation strategies from raw tool outputs, which in this example helped steer the model away from shallow sanitization toward a correct fix. Third, emphasizing security invariants (e.g., enforcing directory containment) instead of isolated syntactic changes can encourage broader, structural repairs that address the root cause of the vulnerability. Finally, organizing feedback into clear, ordered steps may better align with the model’s instruction-following behavior, supporting systematic, multi-location edits. While these observations are drawn from a single case study, they are consistent with the empirical improvements observed for explained feedback in our larger-scale experiments.

\paragraph{\underline{\textbf{Takeaway}}}
Explanation-based repair is most effective when vulnerabilities can be fixed with localized, well-scoped edits, and its benefit diminishes as the required changes become more distributed or configuration-heavy. 
The strongest instruction-followers, GPT-4o and GPT-3.5-turbo, together with DeepSeekV2 on the SecCodePLT and SecurityEval tasks, can translate the explanations into coherent patches, while smaller and most open-weight models often fail to coordinate the necessary updates on CyberSecEval's longer, multi-CWE explanations. This reinforces the broader conclusion that effective post-hoc repair depends not only on understanding the feedback but also on the model’s ability to execute structured, multi-step code transformations.
\section{Discussion}
\label{sec:threats}
\subsection{Threats to Validity}

We acknowledge several threats to the validity of the results and findings of our study.

{\bf The use of CodeQL.} Our evaluation is based on a static vulnerability detection tool, CodeQL, which effectively identifies many common security issues through static analysis, but fails to capture dynamic vulnerabilities or those that manifest under specific runtime conditions. 

{\bf Benchmark scope and representativeness.}
Our study is limited to Python and to three datasets. Python is widely used and offers a practical testbed, but its features may not reflect the security challenges of languages such as C++ or Java, so language-specific vulnerability trends should be studied separately. The three datasets are complementary in their limitations. SecurityEval averages fewer than two samples per CWE, so per-CWE conclusions on it are indicative rather than definitive. {We therefore base per-CWE analysis on SecCodePLT, use SecurityEval for breadth (69 CWE types), and use CyberSecEval---which has SecCodePLT-comparable per-CWE depth but covers only 8 CWE types---for real-world representativeness and trend replication.}
SecCodePLT is synthesized via mutation and may not capture the full complexity of production code. We partially address this by adding the CyberSecEval benchmark, on which our key prevention finding replicates. Even so, no fixed set of benchmarks covers the full and evolving space of vulnerabilities.

{\bf Models studied.}
Our model set is chosen to span code specialization and scale, but it was bounded by compute budget and model availability, and some larger or more recent models were not included for these practical reasons. Our conclusions are therefore drawn over a representative rather than an exhaustive set of LLMs.

{\bf Evaluation protocol.}
We evaluate in a static, single-turn setting, whereas real development is often multi-turn, with developers iteratively refining and debugging code together with an LLM. Our setup may thus underestimate the improvements achievable through conversational, iterative repair. We also fix the decoding hyperparameters (Table~\ref{tab:hyperparams}) for reproducibility and verify that our prevention findings are robust to the number of hints (Figure~\ref{fig:topk_sensitivity}). A systematic sweep over the decoding temperature is left to future work.

{\bf Dependence on a strong feedback model.}
GPT-4o plays three roles in our study: it is one of the evaluated models, the generator of explained feedback, and the LLM-as-a-judge for hint quality. This creates a circularity risk. When GPT-4o repairs its own code using an explanation it wrote, or has its hints judged by itself, its measured performance may benefit from agreement with its own outputs rather than from genuine repair or detection ability, which could inflate its apparent results. We take three steps to limit this risk. First, we report the \emph{direct}-feedback condition, which uses only the raw CodeQL diagnostic and no model-written explanation. We call this condition \emph{oracle-free} because it does not rely on a strong external model acting as a feedback ``oracle'' that supplies the correct explanation, so it measures each model's standalone repair ability with GPT-4o out of the loop. Second, we manually audited 160 GPT-4o explanations and found over 98\% to be correct and actionable, indicating that the explained-feedback gains are not an artifact of self-agreement. Third, the observation of explained feedback being better than direct feedback also holds for the open-weight models, whose explainer (GPT-4o) is a \emph{different} model, so the effect is not specific to GPT-4o judging itself. We do not perform a full human evaluation because manually reviewing every generated program at our scale (thousands of generations across eight models, three datasets, and multiple conditions) is not feasible; we instead rely on CodeQL as a consistent, reproducible detector and validate explanation quality through the targeted 160-sample \emph{human} audit above, in which the authors manually inspected each explanation for technical correctness and actionability. Taken together, these steps make us confident that the explained-feedback effect is genuine rather than a self-agreement artifact.
However, this audit covers only SecCodePLT and SecurityEval, with 10 samples per model per dataset. It does not extend to CyberSecEval, the one benchmark where explained feedback stops outperforming direct feedback on average. Our attribution of that divergence to longer, multi-CWE explanations is therefore a hypothesis rather than a validated conclusion. We cannot rule out that GPT-4o's explanations are simply of lower quality on these real-world, multi-CWE tasks.

{\bf Self-generated hints share the model's blind spots.} In our prevention experiments (RQ1), the model that writes the vulnerability hints is the same model that later consumes them. Weaker models therefore receive weaker hints, and a self-generated hint cannot surface a vulnerability that lies in the model's own blind spots. Our hint-quality analysis of relevance and preciseness (Section~\ref{sec:res_vul_code_gen}), the CWE-definition condition that supplies the target weakness externally, and the self-explained repair condition together bound this effect. Even so, the prevention gains in RQ1 reflect what a model can achieve with its own security knowledge, not an upper bound under independent expert guidance.

\subsection{Suggestions for Developers}
Our findings suggest several strategies for developers aiming to integrate LLMs into secure coding workflows. 
First, 
adapting different LLMs at various stages could be promising: smaller code-optimized models can be used to generate initial code with fewer vulnerabilities, while larger models are applied to vulnerability reasoning and repair. This mixed strategy can offer a cost-effective path to safer code.
Second, it is evident that incorporating fine-grained, explained feedback is essential for effective vulnerability repair, as demonstrated by the significant improvements observed with detailed explanations and actionable suggestions. 
Developers should design their systems to provide rich, actionable insights (e.g., incorporating a powerful reasoning model for vulnerability hints predictions and vulnerability repair suggestions) that guide the LLM in identifying and fixing security flaws.
Third, expanding the set of vulnerabilities is crucial, as the lower performance of LLMs on the complex SecurityEval dataset indicates. By including a broader range of vulnerabilities, developers can enhance model robustness and better address diverse security risks in real-world scenarios.

\section{Conclusion}
In this work, we conducted a comprehensive evaluation  across a diverse set of LLMs for secure code generation and vulnerability repair. Our study revealed that while LLMs are inherently prone to generating insecure code, their security performance can be significantly improved through the incorporation of self-generated vulnerability hints and explained, contextualized feedback. Importantly, the effectiveness of self-generated hints is contingent upon their relevance and preciseness; contextualized feedback that are more relevant and precise can result in lower vulnerability rates in code repair tasks. Beyond these specific results, our study points to a single underlying theme: \emph{recognizing} a vulnerability is not the same as being able to \emph{act} on it. The models that prevent and repair most effectively are precisely those that can operationalize security knowledge---turning a relevant, precise hint or a detailed explanation into a correct, often multi-step edit---whereas weaker models may name the right weakness yet still fail to remove it. This gap between awareness and action is what separates strong from weak models in both prevention and repair, and it is most visible for the multi-step, configuration-level CWEs (e.g., CWE-22 and CWE-611) that resist both. Our conclusions hold within the scope noted in Section~\ref{sec:threats}: a single-turn, Python-only, static-analysis-based setting, a fixed model set, and a dependence on a strong feedback model. These limitations also point to concrete future directions: multi-turn and agentic repair loops that iterate with the detector; self-improving feedback that does not rely on an external teacher model; extension beyond Python and beyond statically detectable flaws; and targeted techniques for the multi-step, configuration-level vulnerabilities that current prevention and repair leave largely unaddressed.

\begin{acks}
The project was sponsored by the Virginia Commonwealth Cyber Initiative (CCI) and the National Science Foundation (Award Number 2311468/2423813). Hao Yan was also partially funded by the GRA Fellowship from the Center for Advancing Human-Machine Partnership (CAHMP) at GMU. The project compute was provided by the Office of Research Computing at George Mason University (https://orc.gmu.edu), which was funded in part by grants from the National Science Foundation (Award Number 2018631).
\end{acks}



\bibliographystyle{ACM-Reference-Format}
\bibliography{custom}










\end{document}